\documentclass{article}

\usepackage[
backend=biber,
style=numeric,
citestyle=numeric,
sorting=none
]{biblatex}

\usepackage[a4paper, total={6in, 8in}]{geometry}

\usepackage{times,amsmath,amsfonts,amssymb,latexsym}
\usepackage{graphicx,epsf}
\usepackage{subfigure}
\usepackage{cancel}
\usepackage{color}

\title{Anisotropic MagnetoMemristance}
\author{
$\ ^1$Francesco Caravelli\footnote{Correspondence: caravelli@lanl.gov}, $\ ^2$ Ezio Iacocca , $\ ^3$Gia-Wei Chern,\\$\ ^1$Cristano Nisoli,$\ ^{4}$Clodoaldo I. L. de Araujo\\\ \\
{\small $\ ^1$ Theoretical Division (T4),
Los Alamos National Laboratory, Los Alamos, New Mexico 87545, USA}\\
{\small $\ ^2$Center for Magnetism and Magnetic Materials,
University of Colorado, Colorado Springs, CO 80918, USA}\\
{\small $\ ^3$Department of Physics, University of Virginia, 
Charlottesville, VA 22904, USA}\\
{\small $\ ^{4}$Departamento de F\`{i}sica, %Laborat\'{o}rio de Spintr\^{o}nica e Nanomagnetismo,
Universidade Federal de Vi\c{c}osa, Vi\c{c}osa,
36570-900, Minas Gerais, Brazil}\\
}

\date{ }

\begin{document}

\maketitle

\begin{abstract}
{In the last decade, nanoscale resistive devices with memory have been the subject of intense study because of their possible use in brain-inspired computing. However, operational endurance is one of the limiting factors in the adoption of such technology. For this reason, we discuss the emergence of current-induced memristance in magnetic materials, known for their durability. We show analytically and numerically that a single ferromagnetic layer can possess GHz memristance, due to a combination of two factors: a current-induced transfer of angular momentum (Zhang-Li torque) and the anisotropic magnetoresistance (AMR). We term the resulting effect the anisotropic magneto-memristance (AMM).  We connect the AMM to the topology of the magnetization state, within a simple model of a 1-dimensional annulus-shaped magnetic layer, confirming the analytical results with micromagnetic simulations for permalloy. 
Our results open a new path towards the realization of single-layer magnetic memristive devices operating at GHz frequencies.}
\end{abstract}

\medskip

\section*{Introduction}
The development of efficient beyond-von Neumann, bio-inspired, or unconventional computing hinges on the realization of novel devices that can be integrated into traditional circuitry. For this reason, memristive devices are an interesting option. Magnetic materials are promising in this regard because of their integration with CMOS and relatively simple device production. 
A \textit{memristive device} is 1-port electrical component that satisfies Ohm's law, $V=R\ I$, and moreover has a dynamical resistance of the form $\frac{dR}{dt}=f(R,I)$~\cite{chua1971,review1,review2}.  An ideal memristor, in which the resistance depends only on the charge $q=\int  I(t) dt$ has not been realized yet. Instead, any generic electrical device that exhibits a pinched hysteresis loop in the current-voltage diagram, whether the memory is stored for long or short term, is currently regarded as a memristive device ~\cite{stru13}. 

From an immediate practical perspective,  memristive devices are of interest because they can be used as a memory. More broadly, they are studied for their collective behavior and applications in neuromorphic engineering \cite{ProbComp5}. For instance, the Strukov-Williams memristor, initially identified while studying titanium dioxide~\cite{stru8}, can be roughly approximated, far from the resistive boundaries, by the functional form~
\begin{equation}
 \frac{dR}{dt}=\beta J-\sigma R+\eta,\quad R_\mathrm{on}\leq  R(t)\leq R_\mathrm{off},
\label{eq:memmod}
\end{equation}
where $J$ is typically the total current density flowing into the device, while $\beta$, $\sigma$, and $\eta$ are constants which need to be determined experimentally. The equation above describes a resistive device whose resistance $R$ is constrained between two values $R_\mathrm{off}>R>R_\mathrm{on}$.  
This is seemingly equivalent to a first-order memristor model with $R(w)=(1-w)R_\mathrm{off}+w R_\mathrm{on}$ and $\frac{dw}{dt}=\alpha w -\beta J$. Note that if $\alpha=0$, this is an ideal memristor as the internal parameter $w$ can be associated to the charge
In general, close to the boundaries $R_\mathrm{off}$ and $R_\mathrm{on}$, eqn.~(\ref{eq:memmod}) loses its validity. Nonetheless, in a memristive device the history of its resistance is not uniquely determined by the applied voltage, but depends also on the device internal state.

The use of memory effects in resistive materials has been suggested for a variety of computing applications ranging from logical gates~\cite{spin1,spin2,spin4,spin5} to unconventional computing~\cite{memr1,memrc1,memrc2}; the long term view is that resistive type of memories can perform a variety of tasks ranging from Boolean computation to machine learning~\cite{ProbComp5,review1,review2}.  Within the context of purely electronic memristive devices, their application has focused on hybrid analog-CMOS devices such as crossbar arrays, in which the memristive effect is exploited as memory devices (switches). These are often used in neuromorphic applications via a field-programmable gate array (FPGA) controller, ranging from neural networks to optimization~\cite{ProbComp5}. 
Such architectures however do not harness the full analog properties of memristive devices, being used as an independent memory unit, with the advantage of being controllable with a single bus. 
There is  a general interest in harnessing the analog power of memristor devices because they exhibit  emergent behavior  in complex interconnected circuits \cite{ProbComp5}. For instance, network of memristors \cite{memrc1,memrc2} can be used for reservoir computing \cite{du} and emergent phenomena such as symmetry breaking and tunneling \cite{Hochstetter2021,caravellitun}.

This work introduces a magnetic memristor which, due to the absence of a tunnel-junction, leads to a smooth response to current, thus making it suitable for analog use. As we discuss in this paper, any soft magnetic material that exhibits anisosotropic magnetoresistance, if the the spin transfer torque is non-negligible, should exhibit a memristance.
The effect is due to the spin-transfer torque~\cite{Beach2008}, and can access the intrinsic magnetization precession frequencies in the GHz regime~\cite{Silva2008,Bonetti2009}.  Having a GHz memristive device implies that these can operate at frequencies compatible with current CPU speeds, and have the potential to be used symbiotically in neuromorphic-von Neumann computer architectures.
%\textbf{We also need to give a good, short motivation for the question: why would it be desirable to have a smooth memrstive behavior at GHz frequencies?}

Different types of memristive devices use magnetic materials and rely on magnetoresistance effects. An initial approach was { magnetic tunnel junctions~\cite{Lequeux2016}, relying on current-induced domain wall motion}. Moreover, such a resistance was shown to depend on differently accessed minor loops, offering a robust multi-level resistive device. More recently, it has been observed that current-induced domain-wall formation in connected Kagom\'{e} artificial spin ices~\cite{amrsi,Chern3,cargwcnis} may give rise to memristive effects. 

The use of magnetic materials as a memristive functional element has led to the development of the rapidly growing field of bio-inspired and unconventional computing. Applications and device design include neuromorphic computing with magnetic tunneling junction oscillators~\cite{grollier1,grollier2}; reservoir computing via domain walls in arrays of magnetic rings inducing ``fading memory''~\cite{allwood}, superparamagnetic ensembles~\cite{Welbourne2021}, and magnetic topological objects~\cite{Prychynenko2018,Zou2021}; and inverse-design or magnonic neural networks~\cite{Wang2021,papp}. {In particular, a hysteretic resistive effect based on the giant-magnetoresistance effect  has been discussed in \cite{munchenberger}.}

These applications, however, do not rely on a memristive effect as defined in Eqn.~\eqref{eq:memmod}. 
%In the GHz regime, there has been also growing interest in spin-torque nano-oscillators for neuromorphic applications \cite{grollier1,Bonetti2009}. 
%\textbf{and why is this a limitation? tie up with previous paragraph}.
%It is an advantage, both for reasons of scalability and for speed of operation the use of fully analog, to obtain smooth resistive switching within 1-port devices \cite{ProbComp5} and fully understand the dependence on the magnetization of the device. On the other hand, smooth memristive effects in magnetic materials and at GHz have been difficult to obtain in analytical form.
On this regard, recent evidence for a memristive effect has been presented for exchange-biased bilayers with mHz field sweeps relying on ``viscous'' dynamics~\cite{Ivanov_arxiv} and for a rather exotic spin-glass heterostructure driven by spin-transfer torque in MHz regime~\cite{Ivanov_arxiv}. % \textbf{[what is this paper?]}
While conforming more closely with Eqn.~\eqref{eq:memmod}, these devices do not achieve GHz memristive effects, thus preventing their integration with current computer architectures.

A na\"{i}ve expectation is that the intrinsic precessional GHz frequencies %and magnetic damping 
of trivial magnetic materials provides a natural environment for memristive effects. Such frequencies can be practically accessed by compensating the magnetic damping with spin-transfer torques~\cite{Ralph2008}, e.g., using spin-torque nano-oscillators~\cite{grollier1}. It is therefore desirable to fully describe memristive effects originating in a single magnetic material due to current. %-induced torques.
This is further advantageous from the point of view of scalability and operation speed of fully analog 1-port devices~\cite{ProbComp5}. 

Here, we analytically demonstrate that a single, trivial ferromagnet exhibits memristive effects at GHz frequencies when an in-plane current both exerts torque via the Zhang-Li mechanism~\cite{Zhang2004} and induces anisotropic magnetoresistance, a combined effect we term {\em anisotropic magnetomemristance} (AMM).
%Here, we theoretically and numerically demonstrate that memristive effects in traditional ferromagnets 
By considering a simple toy model of a {1D} ferromagnetic ring, we {analytically} identify that the memristive effect is enabled by homochiral textures that exhibit a nontrivial topology, akin to %. Our theoretical framework links texture-mediated transport of angular momentum in the form of
spin superfluids~\cite{Konig2001,Iacocca2017,Tserkovnyak2018}. %to dynamic memristive effects.
{This implies that the memristive effect depends on the magnetization state and is, in principle, \emph{reconfigurable}. {Physically, the AMM discussed here can be understood as the inertial dynamics of magnetic solitons and its impact on the magnetoresistance of the material.} We extend our analysis to micromagnetic simulations where we consider a ferromagnetic  {annulus}  with finite thickness. In such a ring, a magnetization state composed of homochiral pairs of domain walls is readily accessible by simple field relaxation, conserving the magnetization's topology, and affording the system with AMM. This allows us to conclude that memristive effects originate from the structural dynamics of the domain walls rather than their current-induced translation. 

{ We summarize here our results. We first develop an analytical framework to study the emergence of AMM in a thin annulus, connecting the current in the device to the voltage drop, and then using the Landau-Lifschitz-Gilbert equation for the dynamics of the magnetization. We then analyze the dynamics of the magnetization in various magnetic states, showing these are stable under certain conditions which we deem ``topological". We then analyze numerically, using micromagnetic simulations, a thick annulus, focusing on common permalloy.}

We expect our results to foster research on analog and reconfigurable memristive devices based on magnetic materials.}
%We verify the memristive behavior by micromagnetic simulations and numerically show that the effect benefits from nonlocal dipole fields by distorting the topological texture in the out-of-plane component.%is impervious to nonlocal dipole fields. \textbf{Need to check if this is true, i.e., if I can numerically excite it without nonlocal dipole}.%with a simple numerical toy model whereby the dynamics are detected via anisotropic magneto-resistance (AMR) and driven by Zhang-Li torque~\cite{Zhang2004}.

\section*{Results}
{Let us now present the main results in the paper. We first develop a theoretical framework to analyze certain magnetic states and formally establish their relationship to resistive and memristive effects. The ideal magnetization states considered in this section allows us to identify the requisites to enable memristance. %We consider ideal states from a theoretical perspective to identify the salient features required to obtain a memristance effect. %We will use this framework to analyze certain magnetic states, some only of theoretical interest. While these states are a little unrealistic from the point of view of real materials, they provide a much necessary intuition to understand the AMM effect, as opposed to the AMR effect.
We will then use this general framework to analyze magnetic states obtained from micromagnetic simulations that take into account nonlocal dipole fields.}% which considers the fact that we have both a finite size  {annulus}  (and thus shape anisotropy) and dipolar interactions. }
\subsection*{Analytical formulation} \label{sec:theor}
We begin by considering the first-order correction to the electric field arising in a soft ferromagnet due to AMR~\cite{amrsi,amr,Chern3}. This correction appears as %If AMR is present, the effective electric field in the material is corrected via 
a combination of current and the magnetization according to \cite{amr,Chern3},
\begin{equation}
\vec{\mathbf{E}}=\rho_0 \vec{\mathbf{J}}+\hat {\mathbf{m}} ( \rho_{||} -\rho_{\perp}) (\hat{ \mathbf{m}}\cdot \vec{\mathbf{J}}),
\label{eq:amr}
\end{equation}
where $\rho_{||}$ and $\rho_{\perp}$ are the resistivity parallel and perpendicular to the normalized magnetization vector $\mathbf{m}$, respectively, and $\vec{\mathbf{J}}$ is the current density. We assume $\rho_{\|}>\rho_{\perp}$. %The physical origin of the effect is well documented, and is due  to the scattering properties of electrons, depending on the orientation of magnetization \cite{amr}. 
The first term in Eqn.~\eqref{eq:amr} is simply Ohm's law in the continuous medium, while the second term is the first order contribution to the electric field due the AMR effect~\cite{amr}. In Ref.~\cite{cargwcnis}, it was shown that it can be recasted as a resistor network with voltage generators in series. Then it is exactly solvable and mapped into an effective memristive effect. While a network analysis is certainly desirable, we focus here on the simple case of a single magnetic element to rigorously derive the memristive effect due to current-induced magnetization dynamics and the magnetization texture. % Such established new behavior opens a new path toward electric memory via magnetic materials. In the present paper, we provide striking theoretical and numerical evidence that magnetic memory can lead to electric memory via this same AMR in an extended 3-dimensional magnetic material. Since a complete analytical understanding of complex continuous magnetic material interacting with currents is lacking, here we focus on a simple model of a magnetic ring, which is also the basis for more complex analysis \cite{allwood}.

\begin{figure}
    \centering
    \includegraphics[scale=0.25]{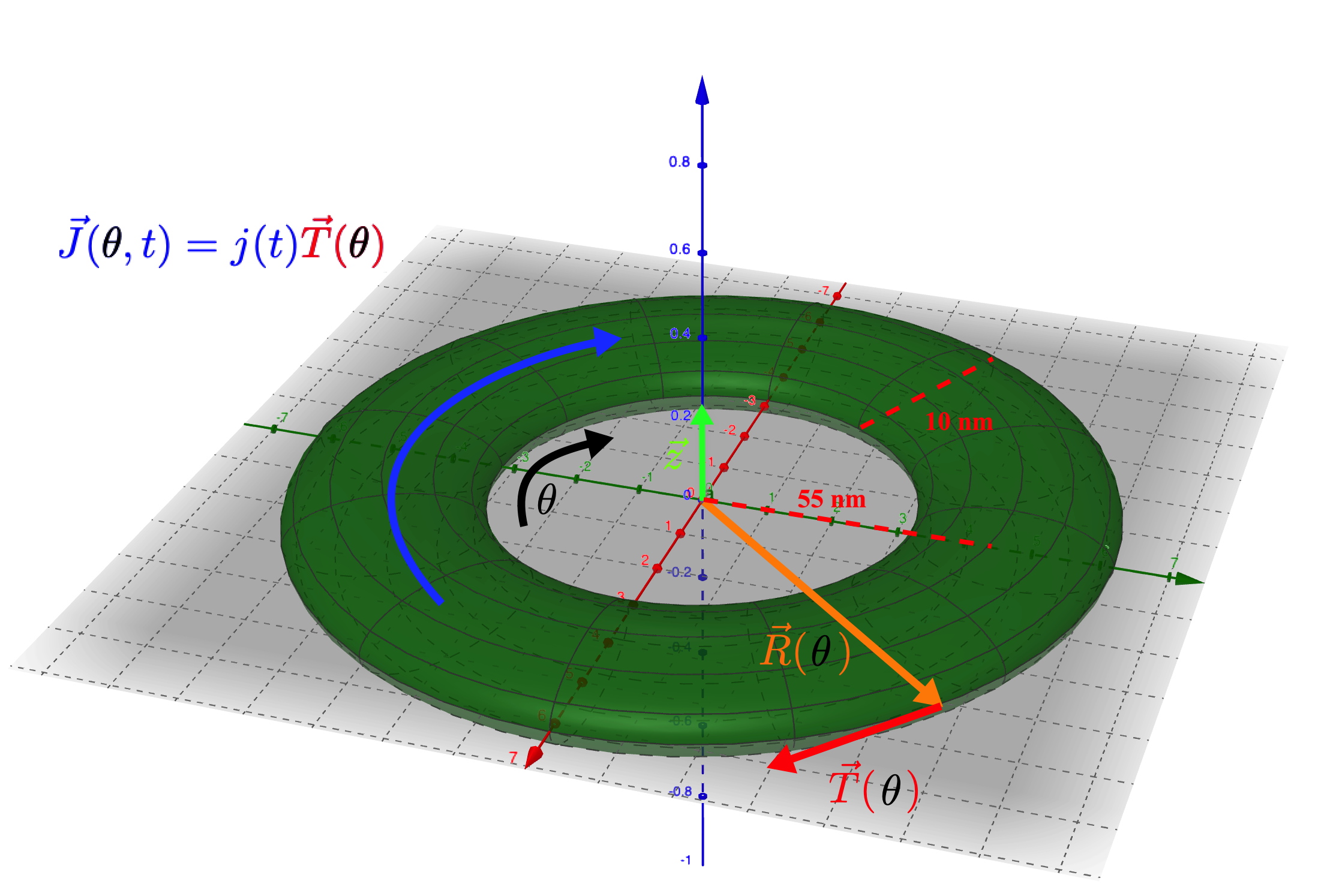}
    \caption{\textbf{The geometry of the annulus under study}.
 We consider a nanoscale magnetic device which has the structure of a thick ring, with the dimensions of $10$nm in thickness and $55$nm radius, and vertical dimension of negligible size but finite. The current configuration, $\vec{ \boldsymbol{J}}(\theta,t)$  runs along the ring, either clockwise or anticlockwise depending on the sign of $j(t)$. %, assuming that a voltage or equivalently a current generator is placed on the ring.
 The  {ring}  coordinate system is given by $\hat{ \boldsymbol{T}}(\theta)$, $\hat{ \boldsymbol{R}}(\theta)$ and $\hat{\boldsymbol{z}}$. In the toy model, the thickness of the  {ring}  is assumed to be negligible, while in numerical simulations finite size effects are considered. }
    \label{fig:ring}
\end{figure}

{While the later simulations will be performed for the case of an annulus geometry, the analytical derivations will be done for the case of a ring, e.g. an annulus of negligible thickness.} The electrical voltage along a certain path $\gamma$ is given by $V(t)=\int_{\gamma} \mathbf{\vec E(\mathbf{\vec J}(t))}\cdot\hat{\mathbf{T}}(\gamma)dl$, where  $\hat{\mathbf{T}}$ is the tangent to the path $\gamma$.  We consider a  {ring}  geometry, whose coordinate system and current physical properties are shown in Figure \ref{fig:ring}. Let us thus consider a closed, quasi-one-dimensional path such as a  {ring}  of radius $r$. In this case, $\gamma$ coincides with the  {ring}  itself, so that $\hat{\mathbf{T}}(\theta)=[-\sin \theta,\cos \theta,0]$ which is a vector oriented clockwise and tangent to the  {ring}  at all points, and where $\theta=0$ corresponds to the positive $\hat x$ axis. Assuming that the current circulates tangent to the path, $\mathbf{J}(t,\theta)=j(t) \hat{\mathbf{T}}(\theta)$, then the effective resistivity is given by $\tilde \rho=V(t)/(2\pi r j(t))$. Analogously and for later convenience, we also define  $\hat{\mathbf{R}}(\theta)$, the in plane vector such that $\hat{\mathbf{R}}(\theta)\cdot \hat{\mathbf{T}}(\theta)=0$, i.e. parallel to the radius of the  {ring}  and directed from the center to the  {ring}  point at $\theta$ (Figure \ref{fig:ring}). In this formulation, we can identify a particular point on the  {ring}  via the angular value $\theta$.%, which in the following we will refer as $\theta$. %\textbf{Ezio: I took a liberty here because resistivity should be in $\Omega$m while the former expression had units of $\Omega$m$^2$}.

%For a magnetic  {annulus}  of radius $r$, the equations can be carried out analytically.
Using Eqn.~\eqref{eq:amr}, we %have an 
obtain the effective resistivity% given by
\begin{eqnarray}
%\tilde R=\frac{V(t)}{j(t)}=R_0+ r \Delta \rho \int_0^{2\pi}  (\hat{\boldsymbol{T}}(\theta)\cdot \hat{\boldsymbol{m}}_t(\theta))^2\ d\theta.
\tilde \rho=\frac{V(t)}{2\pi rj(t)}=\rho_0+ \frac{\Delta \rho}{2\pi} \int_0^{2\pi}  (\hat{\mathbf{T}}(\theta)\cdot \hat{\mathbf{m}}(\theta))^2\ d\theta.\label{eq:res}
\end{eqnarray}
where $\Delta \rho=\rho_{||}-\rho_\perp$ and $\hat{\mathbf{m}}(\theta)$ is the normalized magnetization vector expressed at the point $\theta$ on the ring.  The second term in Equation~\eqref{eq:res}  depends on the magnetic state and originates from the AMR. From it, we see that the resistivity is constrained by  two limiting values, $\rho_\mathrm{off}=\rho_0+\Delta \rho$ and $\rho_\mathrm{on}=\rho_0$, the typical resistivity of the material. These two values are typical for the AMR effect. The value $\rho_\mathrm{off}$ is obtained if $\hat{\mathbf{T}}(\theta)\| \hat{\mathbf{m}}(\theta)$, while $\rho_\mathrm{on}$ if $\hat{\mathbf{T}}(\theta)\perp \hat{\mathbf{m}}(\theta)$.  $(\rho_\mathrm{off}-\rho_\mathrm{on})/\rho_\mathrm{on}=\frac{\Delta \rho }{\rho_0}$. Values for $\Delta \rho/\rho_0$ %\approx 0.03$ 
are documented for a variety of materials~\cite{amr} and it is around 3\% for Permalloy.%, but for Permalloy (Py) we are around 3\% of the change in resistance. 

We are interested in how the effective resistivity of the material changes dynamically. We thus must consider %take into account 
the time evolution of the magnetization vector into Eqn.~\eqref{eq:res}. For a ferromagnet subject to an in-plane current, the magnetization dynamics is described by the Landau-Lifshitz-Gilbert (LLG)~\cite{Gilbert} equation extended with the Zhang-Li torque~\cite{Zhang2004}, given by
\begin{eqnarray}
%\frac{\partial\mathbf{M}}{\partial t}&=&
%-\gamma\mathbf{M}\times\vec{\mathbf{H}}_\text{eff}
%+\frac{\frac{\alpha}{M_s}}{M_\text{S}}\mathbf{M}\times\frac{\partial\mathbf{M}}{\partial t} \nonumber \\
%&+&\frac{pa^3}{2eM_\text{S}}\mathbf{M}\times\big((\mathbf{J}(\mathbf{r})\cdot\nabla)\mathbf{M}\big),
\frac{\partial\hat{\mathbf{m}}}{\partial t}=
-\gamma_g\mu_0\hat{\mathbf{m}}\times\vec{\mathbf{H}}_\text{eff}
+\alpha\hat{\mathbf{m}}\times\frac{\partial\hat{\mathbf{m}}}{\partial t} -\xi \nu\left(\hat{\mathbf{m}}\times(\mathbf{\vec J}(t)\cdot\nabla)\hat{\mathbf{m}}\right)-\nu\hat{\mathbf{m}}\times\left(\hat{\mathbf{m}}\times(\mathbf{\vec J}(t)\cdot\nabla)\hat{\mathbf{m}}\right),\nonumber \\
\label{eq:LLG}
\end{eqnarray}
where $\nu=\frac{p \mu_0}{e M_S (1+\xi^2)}$, and
where we consider the gyromagnetic ratio $\gamma_g$, saturation magnetization $M_\text{S}$,
the vacuum permeability $\mu_0$, Gilbert damping coefficient $\alpha$, polarization ratio of the electric current $p$,  %lattice constant $a$,
electric charge $e$, and the degree of adiabaticity $
\xi$ \cite{Zhang2004}. The effective field $\vec{\mathbf{H}}_\text{eff}$ %, which is composed by %the magnetostatic field,
may include an external magnetic field, magnetocrystalline anisotropy, %Heisenberg
exchange interaction, and nonlocal dipole fields interaction. %It is well known that the Gilbert damping parameter $\frac{\alpha}{M_s}$ depends both on the temperature and the thickness of the film, and thus the size and dimensionality of the material is important. This said, we first aim to show tat LLG equations \ref{eq:LLG} and the AMR effect of eqn. (\ref{eq:amr}) lead to a memristive behavior in a simple example.

%For the case of a  {annulus}  studied here, we map the magnetization dynamics in the polar plane, such that the magnetization vector denoted as $\mathbf{M}_r(\theta)$ lives on a 1-dimensional space and depends only on the polar angle $\theta$ (aside from time). This simplification allows us to carry out an exact analytical derivation.%most of the calculations analytically and will guide our interpretation.

%For the magnetic  {annulus}  we have shown in the {Supplementary Material} B analytically that
As we show {in Supplementary Note 1 and 2}, the combination of both the AMR effect of Eqs.~\eqref{eq:amr} and %the Landau-Lifshitz-Gilbert equation (\ref{eq:LLG})
\eqref{eq:LLG} leads to a time-evolution of the resistivity %of the permalloy
of the form% (see {Supplementary Material}):
\begin{subequations}
\label{eq:memlike_full}
\begin{eqnarray}
%\frac{d \tilde R}{d t}
%&=& \beta(\hat{\boldsymbol{m}}_t) j(t)-  \eta(\hat{\boldsymbol{m}}_t,\vec{\boldsymbol{H}}_{eff}),\nonumber \\
%\beta(\vec{\boldsymbol{M}}_t)
%&=&\frac{pa^3 \Delta \rho}{2eM_\text{S}  }  \int_0^{2\pi} d\theta\  \frac{\partial \vec{\boldsymbol{M}}_t}{\partial \theta}\cdot \Big(\hat{\boldsymbol{T}}\times \vec{\boldsymbol{M}}_t \Big)t_m\nonumber \\
%\eta(\vec{\boldsymbol{M}}_t)&=&r \Delta \rho \int_0^{2\pi}d\theta\  (\gamma \vec{\boldsymbol{H}}_\text{eff}-\frac{\alpha}{M_s} \frac{\partial\vec{\boldsymbol{M}}_t}{\partial t})\cdot \Big(\hat{\boldsymbol{T}}\times \vec{\boldsymbol{M}}_t \Big)t_m\nonumber \\
%t_m&=&\hat{\boldsymbol{T}}\cdot  \vec{\boldsymbol{M}}_t ,
\label{eq:memlike}
\frac{d \tilde \rho}{d t}
&=& \beta(\hat{\mathbf{m}}) j(t)-  \eta(\hat{\mathbf{m}},\vec{\mathbf{H}}_\mathrm{eff}),\\%\nonumber \\
\label{eq:beta}
\beta(\hat{\mathbf{m}})
&=&\beta_a(\hat{\mathbf{m}})+\beta_{na}(\hat{\mathbf{m}})\nonumber \\
\beta_{na}&=&-\frac{\xi \nu \Delta \rho}{2\pi r}  \int_0^{2\pi} d\theta\  \frac{\partial \hat{\mathbf{m}}}{\partial \theta}\cdot \left(\hat{\mathbf{T}}\times \hat{\mathbf{m}} \right)t_m,\\%\nonumber \\
\label{eq:naterm}
\beta_{a}&=& \frac{\nu \Delta \rho}{2\pi r} \int_0^{2\pi} t_m(\theta) r_m(\theta) d\theta \label{eq:aterm}\\
\label{eq:eta}
\eta(\hat{\mathbf{m}})&=&\frac{\Delta \rho}{2\pi} \int_0^{2\pi}d\theta\  (\gamma_g\mu_0 \vec{\mathbf{H}}_\mathrm{eff}-\alpha \frac{\partial\hat{\mathbf{m}}}{\partial t})\cdot \left(\hat{\mathbf{T}}\times \hat{\mathbf{m}} \right)t_m,
\label{eq:aux}\\
t_m(\theta)&=&\hat{\mathbf{T}}(\theta)\cdot  \hat{\mathbf{m}}(\theta) ,\ \ \ \  r_m(\theta)=\hat{\mathbf{R}}(\theta)\cdot  \hat{\mathbf{m}}(\theta) ,
\end{eqnarray}
\end{subequations}
where $\beta(\hat{\mathbf{m}})$ and $\eta(\hat{\mathbf{m}})$ are functions of the magnetization state in the ring. They  indicate how the device reacts to currents and in their absence, respectively. {We stress that these equations were derived with the approximation of negligible thickness, e.g. for a ring.}

Crucially, Eqn.~(\ref{eq:memlike}) takes the form of the Strukov-like memristor, described %as
in Eqn.~(\ref{eq:memmod}) in which $\sigma=0$, while $\beta$ and $\eta$ are not constants but depend on the magnetization state and are thus tunable. In particular, the $\beta$ constant of Eqn.~(\ref{eq:beta}) takes the form of a memristive device whose properties depend on the geometry of the ring, such as the radius $r$, %and
the parameters of the magnetic material, and the magnetization state $\hat{\mathbf{m}}(\theta)$. %details of the magnetic interaction such as the phenomenological parameters $\gamma$, $\frac{\alpha}{M_s}$ for the Landau-Lifschitz-Gilbert equation and $p$ of the Zhang-Li interaction.
From the point of view of the theory of memristors, %the equations 
Eqs.~(\ref{eq:memlike}) represent an infinite-order memristor, i.e., %e.g. 
the device depends on a vector field rather then a single parameter (as for instance the doping depth in oxide-based devices). 

Equations~(\ref{eq:memlike_full}) are the central result of this paper as they rigorously link current-driven magnetization dynamics to memristive effects in magnetic materials subject to AMR effect and Zhang-Li torque. %are useful for a series of reasons.
We then call this combined effect the anisotropic magnetomemristance (AMM).

%Equations~(\ref{eq:memlike_full}) define a memristive effect originating from AMR and current-induced dynamics, we call an anisotropic magnetomemristance (AMM). The Eqs.~\eqref{eq:memlike_full} fully related the AMM to the material's magnetic parameters and are the central result of this paper.

A few comments on the equations above are in order. %It is interesting to note that Eqn.~(\ref{eq:memlike}) takes the form of a Strukov-like memristor described %as
%in Eqn.~(\ref{eq:memmod}) in which $\sigma=0$, while $\beta$ and $\eta$ are not constants but depend on the magnetization state and are thus tunable. %dynamical). T
First, the quantities $\beta_{na}$ and $\beta_{a}$ are the magnetization-dependent non-adiabatic and adiabatic coefficients which, if non-zero, induce a memristive effect. %This effect is tunable only by intrinsic material parameters encapsulated by $\nu$, such as the polarization $p$ and the saturation magnetization $M_s$. %change in the %dependence on the 
%resistivity due to a %change on the 
%current density.
{Another interesting feature is that} the smaller the  {annulus}  radius, $r$, the larger the $\beta_{a}$ and $\beta_{na}$ coefficients are; on the contrary, $\eta$ does not depend on the  {annulus}  radius. This observation implies %, which implies that depending on the size of the  {annulus}  one can have 
a crossover between two regimes: a memristive behavior for small radii %current-dominated change in resistivity, 
and a current-independent resistivity change due to the $\eta$. In addition, the parameter $\eta$ contains the balance between the conservative and dissipative terms of the LLG equation. This implies that memristive behavior is primarily due to the magnetization texture present in the material.

{In addition to the technical comments above, let us briefly provide here a physical interpretation of the equations. The $\beta$ term is the term that leads to a memristive effect, as the resistance changes as a function of the current history, and can lead to a pinched hysteresis loop in the current-voltage diagram. Since this term has two components, originating from the adiabatic and non-adiabatic terms in LLG, both contribute to memristance. %However, the non-adiabatic term will be typically the one dominating in most cases. 
Because of the vectorial dependence of both these terms, the direction of the magnetization with respect to the plane spanned by the  {annulus}  is of key importance. In order for a memristance to be present, the magnetization will have to not lay on the plane, and in particular revolve around the ring, originating from the $\hat{\mathbf{T(\theta)}}\times \hat{\mathbf{m}}$ dependence. %\textbf{Ezio: I am not sure I understand the sentences below. It is jumping to too many conclusions. Can we just mention what is the relevant ``dynamics'' here? Let's also avoid spin superfluids here.} See now
In addition to this, the magnetization will also have to be dynamical in a non-trivial manner. For instance, domain walls that can annihilate in the  {annulus}  can lead only to a transient memristance, but not a stable one. {While these conditions appear to be restrictive, we find that physically achievable} magnetic states that cannot be easily removed via spin transfer torque lead to a more stable memristance effect. This is one of the key features of the  {annulus}  geometry, which naturally can induce states that are ``topological". We will see below that magnetic states that we find to lead to a noticeable memristive states are imperfect, e.g. arising from the interplay of the magnetodynamics of the LLG equation with non-trivial magnetic states.}
%While $\eta$ depends only on the gyromagnetic ratio and the damping constant $\alpha$, $\beta$ depends linearly on the the parameter $\eta$ which is inversely proportional to the saturation magnetization and directly proportional to the polarization ratio $p$. Clearly, both $\eta$ and $\beta$ are zero if $t_m=0$, and thus magnetizations perpendicular to the plane of the  {annulus}  lead, at the same time, to no extra resistivity to AMR (since $\rho=\rho_0+\frac{\Delta \rho}{2\pi} \int t_m^2 d\theta$, nor to memristive effects, since the time derivative is zero.

Naively, one would expect an ideal memristor would be obtained if $\eta=0$ for any magnetization state and time. However, we must recall that $\beta$ is not a constant, and thus $\eta=0$ does not imply that the internal parameter can be directly be associated to the charge. 
Moreover, $\eta=0$ is in general unlikely by pure dissipative action on the conservative term since the effective field is nonlocal. However, we will discuss a particular magnetic state in which this is possible.
The geometrical terms in the integrand provide other possibilities to nullify $\eta$. If $t_m=0$, which implies a magnetization state perpendicular to the ring's circumference, i.e., pointing radially or fully out-of-plane. The latter state could be easily achieved by a saturating magnetic field applied normal to the plane. However, this condition will also nullify $\beta$ and the memristive effect would be suppressed. If $\hat{\mathbf{T}}(\theta)\times  \hat{\mathbf{m}}=0$ then the only allowed magnetization state is the ground state, i.e., a magnetization parallel to the ring's circumference. This case also imply that $r_m(\theta)=0$ and again $\beta=0$ for all cases. This analysis demonstrates that a magnetic  {annulus}  cannot be an ideal memristor.

%pattern of the material which lead to a pinched hysteresis loop (e.g. a memristor).

%Following the parallel with eqn. (\ref{eq:memmod}), the fact that $\sigma=0$ and $\beta,\eta$ are non-trivial functions of the magnetization imply a rather complex dynamical behavior of the effective resistance on the internal magnetization state. In particular, Eqn.~(\ref{eq:memlike}) takes the form of a memristive device whose properties depend on the geometry of the ring, such as the radius $r$, %and
%the parameters of the magnetic material, and the magnetization state $\hat{\mathbf{m}}(\theta)$. %details of the magnetic interaction such as the phenomenological parameters $\gamma$, $\frac{\alpha}{M_s}$ for the Landau-Lifschitz-Gilbert equation and $p$ of the Zhang-Li interaction.
%From the point of view of the theory of memristors, %the equations 
%Eqs.~(\ref{eq:memlike}) represent an infinite-order memristor, i.e., %e.g. 
%the device depends on a vector field rather then a single parameter.  

%Equations~(\ref{eq:memlike_full}) are the main finding of this letter as they rigorously link current-driven magnetization dynamics to memristive effects in the magnetic materials subject to AMR effect and Zhang-Li coupling. %are useful for a series of reasons.
%We then call such behavior the Anisotropic MagnetoMemristive effect (AMM).

\subsection*{Magnetization states, resistivity and memristance} \label{sec:anal}

%The magnetization-dependent quantities $\beta(\hat{\mathbf{m}}(\theta))$ and $\eta(\hat{\mathbf{m}}(\theta))$ indicate %are interpreted as 
%how the device reacts to currents and in their absence, respectively. %(from the point of the effective resistivity) to currents (and the absence thereof). 
%From this toy model we can extract 
We now analyze how certain magnetization patterns could affect the resistance and memristance in this toy model. 
First, let us note that the resistance depends on $t_m(\theta)$, as $\rho=\rho_0+\frac{ \Delta \rho}{2\pi}\int t_m(\theta)^2 d\theta$ and that there are magnetic states that would lead to no {change of} resistance and no memristance. We emphasize that the memristive effect is purely due to the interplay between AMR and spin-transfer torque.

Here we focus on the non-adiabatic term, which is larger in real materials. However, an analysis of which states contribute to memristance for the adiabatic term of eqn. (\ref{eq:naterm}) is provided in {Supplementary Notes 1, 4 and 5 of the Supplementary Material.} 
In the case of the non-adiabatic term in eqn. (\ref{eq:aterm}), the correction depends on $r_m(\theta)$ and $t_m(\theta)$, which means that the the magnetization must not be perpendicular to both the radius and tangent to the  {annulus surface}  at all times.

Thus, a memristive behavior is nullified %The first thing worth noticing is that 
if the magnetization is everywhere \textit{parallel} to the current or, equivalently, the ring's circumference. % in this case), the device is not excitable. This is because both $\beta(\hat{\mathbf{m}}(\theta))=0$ and $\eta(\hat{\mathbf{m}}(\theta))=0$ due to their explicit dependence %depend explicitly
%on $\hat{\mathbf{T}}\times \mathbf{m}$ and $r_m(\theta)$.
A similar fate befalls onto magnetization states where %also occurs if
$\partial_\theta \hat{\mathbf{m}}=0$, i.e., %e.g. if
when the the device is statically magnetized in a particular direction. This means that memristive effects can only happen as a function of current %such device behaves ``memristively" 
if the magnetization texture along the  {annulus}  %in the device
is non-trivial. %Another important comment is how the device behaves with respect to the application of an external field. 

%---
The typical magnetodynamics scale set by the gyromagnetic ratio and the effective magnetic field is contained in the term $\eta(\theta)$ which produces a change in the device's resistivity in non-equilibrium independent of the current density. In a steady state, the damping will exactly cancel out any dynamic contribution and $\eta(\theta)=0$ so that the device will have a finite resistivity. %For the case of the ring, 
In addition, the effective magnetic field %only in the material
is given by $\vec{\mathbf{H}}_\text{eff}=\vec{\mathbf{H}}_0+\lambda\partial^2_{\theta\theta}\hat{\mathbf{m}}-H_{k}(\hat{\mathbf{m}}\cdot\mathbf{\hat T}(\theta))\mathbf{\hat T}(\theta)$. Consequently, the main factor determining the memristance is the static magnetization state in the ring. 

In addition to trivial magnetization states, it is theoretically possible to have textures in our toy model. Non-trivial magnetization states in rings have been theoretically suggested as means to store energy in the form of angular momentum~\cite{Tserkovnyak2018}. Such a state exhibits a smooth, continuous rotation of the magnetization about its perpendicular-to-plane axis and is formally analogous to mass superfluidity or superconductivity~\cite{Konig2001}. Indeed, a {n annulus}  provides a suitable geometry to realize periodic boundary conditions that, in principle, would stabilize a spin superflow with a quantized number of periods $n$ in the absence of dipolar fields~\cite{Iacocca2017}.

We now analyze the predicted resistivity and memristance of a variety of possible magnetization states.

\subsubsection*{Fully magnetized states.} Since both $\eta$ and $\beta$ depend on $t_m(\theta)$ and $r_m(\theta)$ which are the projection on the tangent and radial vectors of the ring, any out-of-plane or uniformly in-plane magnetization will lead to a pure resistive state. This can be easily achieved, e.g, by an external field saturating the magnetization.

%\textit{Ground state.} Certain states do posses {AMR}, but no memristance. An example is a state such that $\hat{\mathbf{m}}(\theta)\propto \hat{\mathbf{T}}(\theta)$, i.e., the magnetization is curled around the ring, i.e. the ground state {also known as the vortex state~\cite{exp2,exp4,exp5}}. In fact, while this magnetic state contributes to the resistivity, it is not hard to see that $\beta_a$ is zero as $r_m(\theta)$ is zero everywhere;  $\beta_{na}$ is alzo zero as $\hat{\mathbf{T}}(\theta)\times \hat{\mathbf{T}}(\theta)=0$. Thus, the ground state has no memristance, as it is confirmed below via micromagnetics of extensive rings.

In real materials, if the sample is prepared in a state in which $\mathbf{m}$ has most of its components in %on
the plane, then the effect of an external field ${\mathbf{H}}_0$ produces %on the plane has 
only a small correction to the resistivity. %resistance.
We found {(see Supplementary Note 3 of the Supplementary Material)} that the correction is of the same order of the out-of-plane component of $\mathbf{m}$, implying that the sample is robust to in-plane external field perturbations.

{\textit{Topological states.}} %Spin superflow and $2\pi$ rotations.} To describe what, in analogy to spin superfluids~\cite{Konig2001,Iacocca2017,Tserkovnyak2018} we will call the spin superflow, 
{ As we have seen, certain natural states on the  {annulus}  do not support a memristive effect.}
{Let us now attempt at introducing a set of states with full rotations of the magnetization.} %kinks, which we will define soon. } 
We introduce a generic magnetization texture of the form
%One important comment we wish to make here is that there some special magnetic states. These are given by stable magnetic states compatible with the symmetry of the ring, and which protect the resistance state, e.g. the memristance effect does not occur if the time dependence leads only to a shift in the $\theta$ directions. These states are pointed out in the {Supplementary material}. An example is given by helical states.
%If we assume that $||\mathbf{M}_t||=m_0$ and $\vec{\boldsymbol{M}}_t\cdot \hat{\boldsymbol{T}}=0$, it is easy to see that these are state of the form
%\begin{eqnarray}
%\mathbf{m}(\theta)=(M_x,M_y,M_z)&=& m_0\Big(\sin(\phi)\cos ( \theta),\nonumber \\
%& & \sin( \phi) \sin(\theta),\nonumber \\
%& & \cos(\phi) \Big)
%\end{eqnarray}
$\hat{\mathbf{m}}(\theta)=\left(\sin(\phi)\cos (n \theta), \sin( \phi) \sin(n\theta),\cos(\phi) \right)$,
%which we call helical states, and 
where $\phi(\theta)$ can be an arbitrary function that represents the out-of-plane component of the texture {and $n$ quantifies the topology of the state}. {When $\phi(\theta)=\pi/2$, the magnetization is fully in-plane and the parameter $n$ defines the number of rotations around the ring. For $n=1$, we recover the ground or vortex state discussed above. For $n\geq2$, such an equation describes a smooth, coherent rotation of the magnetization known as a spin superflow %}A perfect spin superflow corresponds to $\phi(\theta)=\pi/2$ while a helical state to %. In order to have $2\omega $ wrappings around the ring, we need to define 
%$\phi=\omega \theta$.
%{In fact, when $\phi=\pi/2$ the state is known as a spin superfluid 
stemming from its formal analogy to mass superfluidity~\cite{Konig2001}. In these states, the homochiral phase rotations along $\theta$ lead to a net pure spin current which has lead to proposed applications in storage \cite{Tserkovnyak2018} and long-distance spin transport \cite{Sonin2010, Takei2014, Iacocca2017,Iacocca2019b, Yuan2018, Stepanov2018}}.
{These states are interesting because they should be in principle dynamical and non-easily removable via the spin transfer torque. %In magnetism, these have been studied in the literature since their introduction \cite{Konig2001} and experimental realization \cite{Yuan2018}.}
%{
Unfortunately, from Eqn.~\eqref{eq:beta}, a spin superflow leads to $\beta(\theta)=0$. Another state of interest is obtained when $\phi=\omega \theta$, which we call a helical state. Helical states} preclude any AMR change, i.e., %is not hard to see from eqn. (\ref{eq:res}) that these are states for which 
$\tilde R=R_0$ from Eqn.~\eqref{eq:res}. %, e.g. the AMR does affect the resistance.
%The helical state is shown in Figure \ref{fig:tops} for $\omega=3$.
Therefore, neither pure spin superflows nor perfect ``helical states" produce memristive effects ({see Supplementary Note 5 of the Supplementary Material}).

{This theoretical analysis allows us to state that only magnetic states with both topological and ``helical'' characters exhibit memristive effects. This implies that non-trivial states must be artificially created in magnetic materials to support memristive effects. Such non-trivial states generally involve one or more pairs of homochiral domain walls that have been discussed in the context of spin superfluids in nanowires~\cite{Iacocca2017,Sonin2010,Takei2014,Iacocca2019b} or in nanorings~\cite{exp1,exp3} as we discuss below.}

\subsubsection*{Magnetic states dominated by shape anisotropy.} Here we would like to discuss how the interplay between long-range interactions and topology can lead to non-trivial memristive relevant states. In realistic magnetic materials, often the long-range dipolar interaction is the dominating force, leading to the so-called shape anisotropy. Combined with the exchange interaction which favors parallel spins, there are two degenerate ground states on a magnetic  {annulus}  with magnetization $\hat{\mathbf m}(\theta) = \pm\hat{\mathbf T}(\theta)$. The magnetization is curled around the  {annulus}  either clockwise or counterclockwise. Such minimum energy states are also known as the vortex state~\cite{exp2,exp4,exp5}. The vortex state is an example of magnetic state which possesses AMR, but no memristance. In fact, while this magnetic state contributes to the resistivity, it is not hard to see that $\beta_a$ is zero as $r_m(\theta)$ is zero everywhere;  $\beta_{na}$ is alzo zero as $\hat{\mathbf{T}}(\theta)\times \hat{\mathbf{T}}(\theta)=0$. Thus, the ground state has no memristance, as it is confirmed below via micromagnetics of extensive rings. {We numerically find, however, that these states exhibit a small memristive effect when excited with a strong current. This small effect arises from the periodic modification of the ground state that is not captured with the formalism presented so far. A theoretical analysis of this case is presented {in Supplementary Note 4 of the Supplementary Material.}}

{The magnetic texture that interpolates the two different vortex states with opposite circulation is called the domain wall (DW). Since DWs have to be created and destroyed in pairs on a ring, they are also topological defects on top of the minimum energy state. In fact, each DW can be viewed as a mesoscopic object that carries a net magnetic charges: positive charge for a tail-to-tail DW, and negative charge for head-to-head DW. Metastable states on the  {annulus}  can thus be classified according to the number $n_{\rm DW} = 2, 4, \cdots$ of DWs. Importantly, DWs play a crucial part in the memristive effect of a magnetic  {annulus}  by allowing for a non-zero non-adiabatic $\beta_{na}$ term. Since the magnetization $\hat{\mathbf m}(\theta)$ changes from $+\hat{\mathbf T(\theta)}$ to $-\hat{\mathbf T}(\theta)$ in the vicinity of a DW, thus allowing for a finite gradient $\partial \hat{\mathbf m}/\partial \theta$. Moreover, magnetization is forced to deviate from the tangent direction $\pm \hat{\mathbf T}(\theta)$, giving rise to a nonzero $\hat{\mathbf T}\times \hat{\mathbf m}$ as well. As indicated in Eqn.~(\ref{eq:naterm}), these are the two ingredients for a nonzero $\beta_{na}$ term. {As it will be demonstrated in the next Section, the two domain-wall states (2DW) indeed exhibits a significant memristance.  Experimentally, these states have been observed in magnetic nanorings~\cite{exp2,exp4,exp5}. Numerically, such a pair of domain walls can be readily stabilized by including nonlocal dipole fields.} }

%\hline
%\textbf{This seems to be true from simulations, but we need to formally show it. It would be also good to have a figure that illustrates this somehow.}

%For a finite memristive effect, it directly follows from Eqn.~\eqref{eq:memlike} that the resistivity is scaled by a factor $\beta(\theta) j_0/\Omega$, where $j(t)=j_0\cos{(\Omega t})$. Reasonable parameters for Py \textbf{(check)} are $a=0.3$~nm, $\Delta \rho=0.1$~$\Omega$m, $p=0.6$ and a  {annulus}  radius of $100$~nm, the prefactor of $\beta(\theta)$ is $50k$~$\mu\Omega$m$^3$/C, where $k$ is the quantization number for the non-trivial magnetic state. This implies that a current density of $j_0=10^{10}$~A/m$^2$ and $\Omega=1$~GHz leads to a change in resistivity of $500k$~$\mu\Omega$m or a change or resistance of $800k$~$\Omega$ along the ring. \textbf{Can we compare with something?}

%\begin{figure}
%    \centering
    %\includegraphics[scale=0.3]{FiguresRing/Neel3.png}\\
%    \includegraphics[scale=0.1]{Helical.png}\\
%    \caption{choose which you like best}
%    \label{fig:tops}
%\end{figure}

\subsection*{Simulations for magnetic annulus} \label{sec:sims}

The analytical treatment %we
described above provides a useful guide into the phenomenology, but it%has some important limitations. First of all, any realistic material will have a non-zero thickness and width. 
is limited to an ideal one-dimensional case. In reality, a finite cross-section of a  {annulus}  would produce a non-local dipole field that in turn modifies both the stabilized magnetization states and the dynamics. %This is important for a variety of reasons, including the fact surface effects can give important contributions to the dynamics of the magnetization. In order to provide further evidence that the theory of the memristor  {annulus}  is not far off from realistic extended structure, we have performed simulations of an extended  {annulus}  with non-negligible thickness in Figure \ref{fig:ring}. The size of the  {annulus}  is shown in Figure \ref{fig:ring} (top figure), where we see that the thickness of the object is $20$nm, while the total linear size of the  {annulus}  is approximately $120$nm.
To account for the non-local dipole contribution, we perform micromagnetic simulations with use of the open-source package MuMax3~\cite{Vansteenkiste2014} (details are provided in the Methods section and Supplementary Note 6 of the Supplementary Materials). 

We simulate a Py  {annulus}  with a mean radius of $55$~nm, a  {annulus}  width of $10$~nm, and thickness $10$~nm. An analysis of the effects of the dimensions  of the annulus is provided in Supplementary Note 7 of the Supplementary Materials.%We used the software MuMax with 
We use standard material parameters $M_s=790$~kA/m, $A=10$~pJ/m, and $\alpha=0.01$. The Zhang-Li torque polarization parameter is set to $0.56$ and we also include a small non-adiabatic term $\xi=0.1$, and for permalloy we have $\rho_0=123~ \Omega$~nm.

\begin{figure}
    \centering
    \includegraphics[scale=0.4]{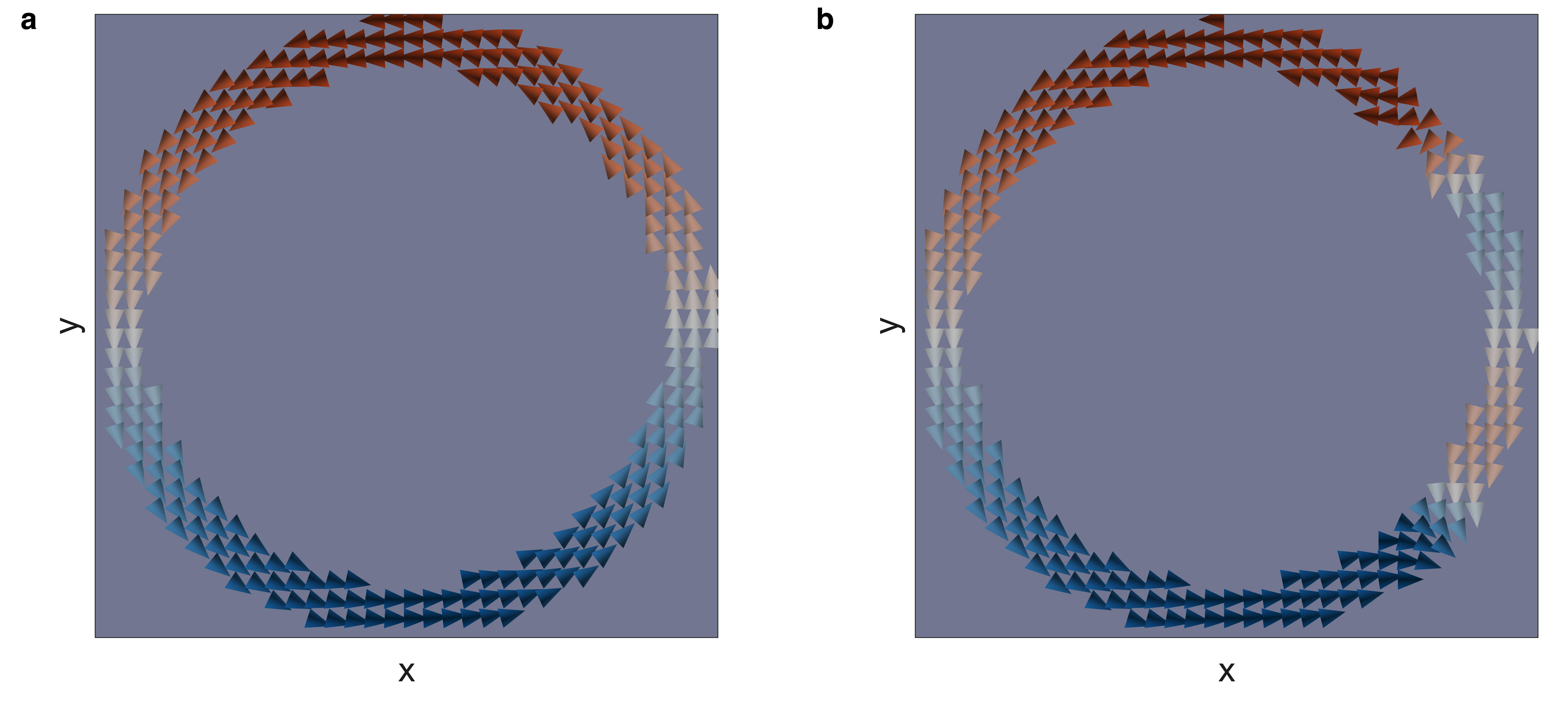}
    \caption{\textbf{The magnetic configurations.} Stabilized \textbf{a} ground state and \textbf{b} {2DW (domain wall)} state utilizing a tiered energy minimization. {The in-plane magnetization component is shown by cones and colored by the $m_x$ value.}}
    \label{fig:states}
\end{figure}

%{add here something about simulations, ground state, in plane, out of plane,  then refer to methods}
We consider four initial magnetic configurations. In all cases, we follow the relaxation protocol described in the Methods to quench spurious dynamics. The ground state is the magnetization along the ring, stabilized by shape anisotropy, as shown in Figure \ref{fig:states} \textbf{a} obtained numerically. In-plane and out-of-plane magnetized states are achieved applying an external magnetic field of magnitude 1~T along the $x$ and $z$ directions, respectively. Finally, we consider {homochiral domain walls stabilized by initializing the magnetization with a finite number of topological defect, %a mixed spin-superfluid state (MSS) which is initialized 
%according to a perfect spin superflow of the form $\hat{\mathbf{m}}(\theta)=\left(\sin(\phi)\cos (n \theta), \sin( \phi) \sin(n\theta),\cos(\phi) \right)$, {%for a given red 
and minimizing the energy. The case of two domain walls is} %,  as 
shown in Figure \ref{fig:states} \textbf{b}. %, obtained numerically.
%In all cases, we follow the relaxation protocol described in the Methods to quench spurious dynamics. 
The final magnetic state is a localized kink of the form $
\hat{\boldsymbol{m}}(\theta)=a(\theta) \hat{\boldsymbol{T}}(\theta)+b(\theta) \hat{\boldsymbol{K}}(\theta)$
where we assume $\hat{\boldsymbol{K}}\cdot \hat{\boldsymbol{T}}=0$, and thus $|\hat{\boldsymbol{m}}|^2=a(\theta)^2+b(\theta)^2=1$. 
%The state is such that we have a dipole induced ground state for most of the ring, with a localized magnetic kink.
We have shown in {Supplementary Note 5 of the Supplementary Material} that such a magnetic state can have both AMR induced resistance and memristance. {This type of states have been obtained seen both numerically and experimentally in the past \cite{exp1,exp2,exp3,exp4}. These states are topologically distinct from the ground state~\cite{braun2012}. %, are topological. 
%\textbf{(Ezio: I have no preference for the sentence to follow. If we think this could open a pandora's box on the definition of the winding number, better to remove it.)} This is due to the fact that these states have a non-trivial winding number, defined by the number of times the magnetization wraps around the ring.
}

We analyze the simulation results by computing the change in voltage and the change in resistivity for each state upon the application of a time-varying current density of magnitude $j$ and frequency $f$. In the following, we discuss variations in voltage, deviating from the resistive state. These can be studied via
 $\Delta V= (R-\langle R\rangle)j$, with $\langle R\rangle$ being the time average resistance. We see that if $R$ is constant, then $\Delta V=0$, which is associated to a pure resistance.

\begin{figure}
    \centering
    \includegraphics[scale=.65]{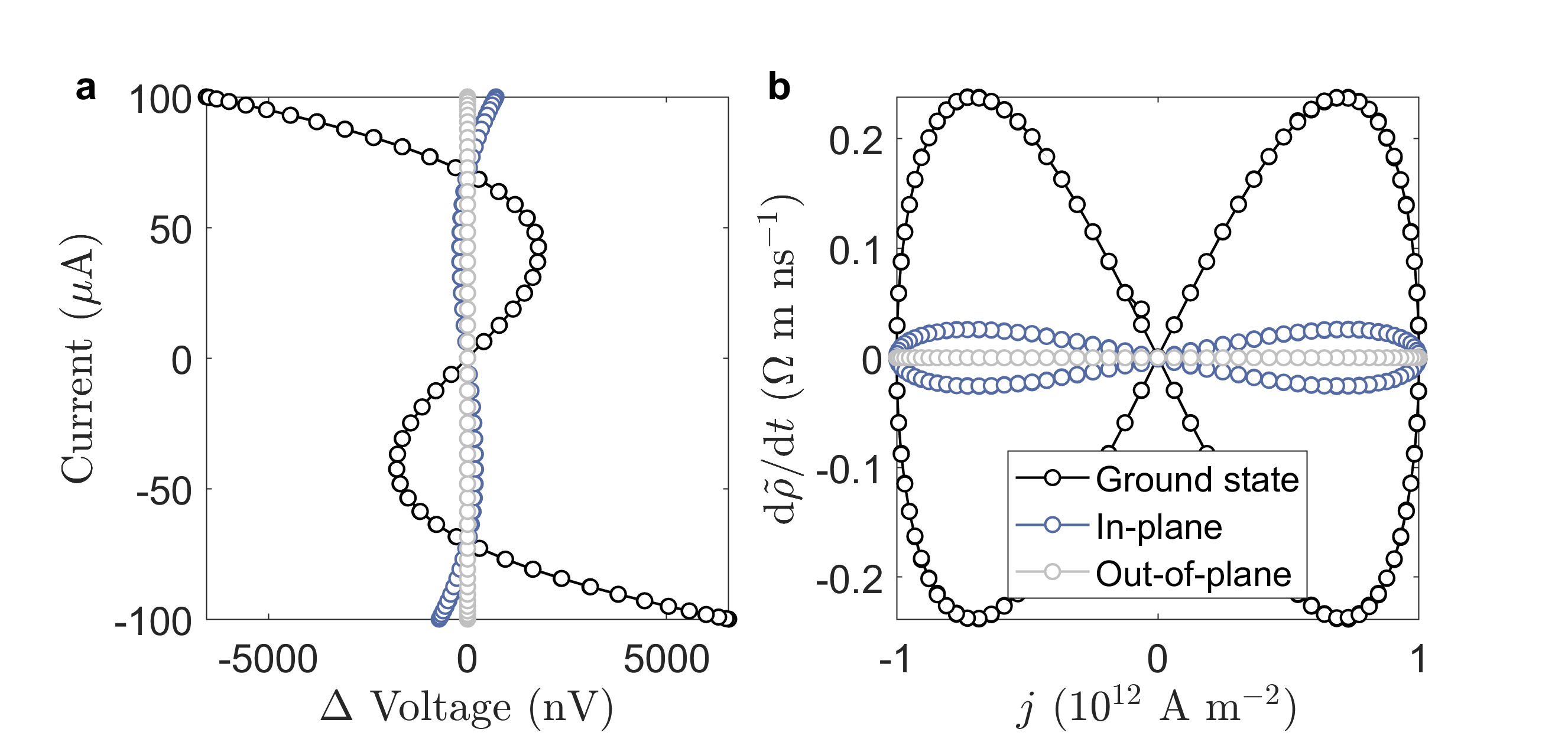}
    \caption{\textbf{Hysteresis curves.} Change in the resistivity $\rho$ of the ground state (black curve), the in-plane magnetization state (blue) and the out-of-plane magnetic state (light grey). \textbf{a} We plot the current-voltage (I-$\Delta V$) diagram of the resistive device as a function of the magnetization. We can see that for currents up to a 100 $\mu$A the resistive states do not possess a noticeable hysteresis in the nanovolts. Yet, the anisotropic magnetoresistance (AMR) leads to a nonlinear resistance state. \textbf{b} We plot the change in resistivity as a function of the current density $j$, which is in units of $10^{11}$~A~m$^{-2}$. From this plot we can see that the ground state and the in-plane do in fact have a memristance effect, which is however very small. For the out-of-plane state instead, the memristance is zero, as predicted by our model. }
    \label{fig:sims}
\end{figure}
For the ground state, in-plane, and out-of-plane magnetization states, the change in voltage versus the current density in a cycle of frequency $f=1$~GHz is shown in Figure \ref{fig:sims}. For these simulations a rather large current density of $10^{12}$~A/m$^2$ was used to discern effects above numerical noise. The $I-V$ curves are shown in Figure \ref{fig:sims}(a), where we see that a memristance effect is not noticeable at a $nV$ scale. This is consistent with the predictions of the toy model.

\begin{figure}
    \centering
    \includegraphics[scale=.2]{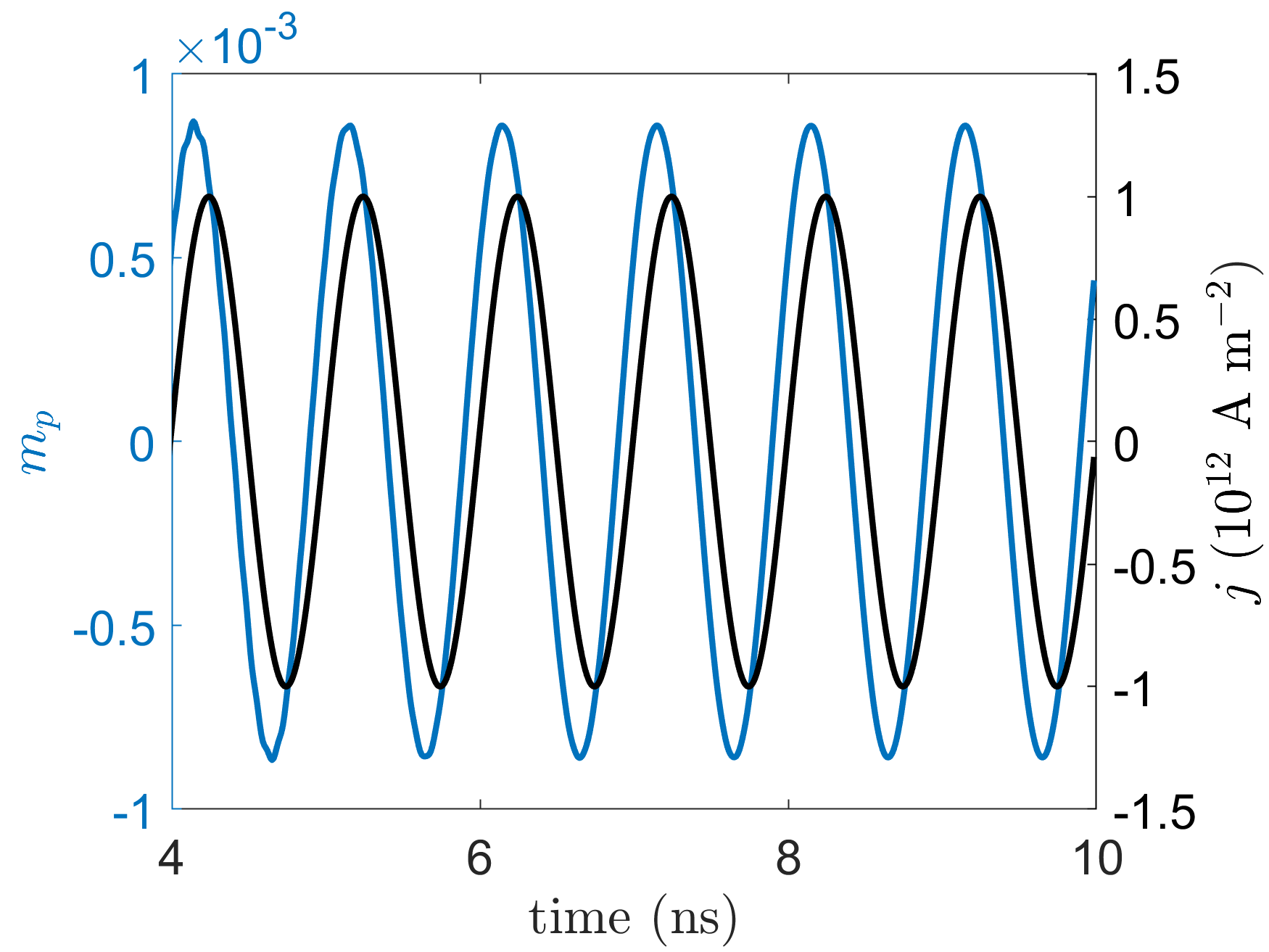}
    \caption{\textbf{Averaged magnetization.} {Spatially averaged magnetization dynamics along the path $T(\theta)$ from micromagnetic simulations of the ground state subject to a microwave current of 1~GHz and amplitude $10^{12}$~A~m$^{-2}$. The magnetization component $m_p$ is perpendicular to the path $T(\theta)$ is shown by a blue curve and the microwave density current $j$ is shown by a black curve. }}
    \label{fig:path1}
\end{figure}

 However, because the thickness of the  {annulus}  gives a small memristance effect that can be seen when looking at $\frac{d\rho}{dt}$.
First,  $\frac{d\rho}{dt}=0$ corresponds to pure resistance. That is the case of  out-of-plane magnetization. %, which is consistent with the theoretical results from the previous section with the  {annulus}  toy model. 
We should expect this property to be true also in real materials with a finite thickness. For the cases of the in-plane and the ground state we note that a small change in resistivity is indeed present, and depends on the current. While a memristance in the ground state and in-plane magnetized state appears to be at odds with the symmetry arguments presented above, we note that the strong current density used in this simulation induces a large enough magnetization dynamics. %This, in turn, produces a memristive effect based on $\beta_a$, i.e., the states presented in Fig.~\ref{fig:nonadmag}.
{These dynamics compose a periodic modification that gives rise to a memristive effect, as elaborated in {Supplementary Note 4 of the Supplementary Material} when discussing the effect of the adiabatic torque component. Figure~\ref{fig:path1} illustrates these dynamics by plotting the time-evolution of the average magnetization perpendicular to the path $T(\theta)$ (see method for its implementation). The periodicity of the dynamics corresponds to the injection current, demonstrating that this memristive effect originates from time-lagged, current-induced dynamics.}

The curves we observe for $\frac{d\rho}{dt}$ are Lissajous curves. Intuitively, this is due to the fact that while the magnetic ground state follows the geometry of the ring, the non-adiabatic coupling leads to a rotation of the magnetization vector. {As discussed in Supplementary Note 5 of the Supplementary Material, we associate this memristance to an adiabatic effect only accessible at large current densities.}%, which then leads to a small shift between the non-adiabatic maximum state and the non-adiabatic minimum state, as shown in Figure \ref{fig:nonadmag}. 
We wish to stress that the fact that we observe a non-zero area associated with these curves is indicative of a memristive effect which depends on an internal state. The pinched hysteresis loop in the $\frac{d\rho}{dt}$ vs $j$ diagram implies a zero decay constant $\eta$. Indeed, if the change in resistivity was proportional to the current density $j$, we should expect to see a line in these plots, with a slope given by a constant $\beta$ parameter. The fact that we see a Lissajous curve means that the parameter $\beta$ is not constant in time. In particular, the periodicity of $\beta$ is exactly the one of the driving $j$. 

%\textbf{revise below}

 %\textbf{non-adiabatic or adiabatic?}. Smaller current densities lead to a change in resistivity on the order of numerical noise. Therefore, we conclude that this large-current-density effect may be visible only for simulations where Joule heating and the progressive damage of a material due to current is not modeled.

\begin{figure}
    \centering
    \includegraphics[scale=.5]{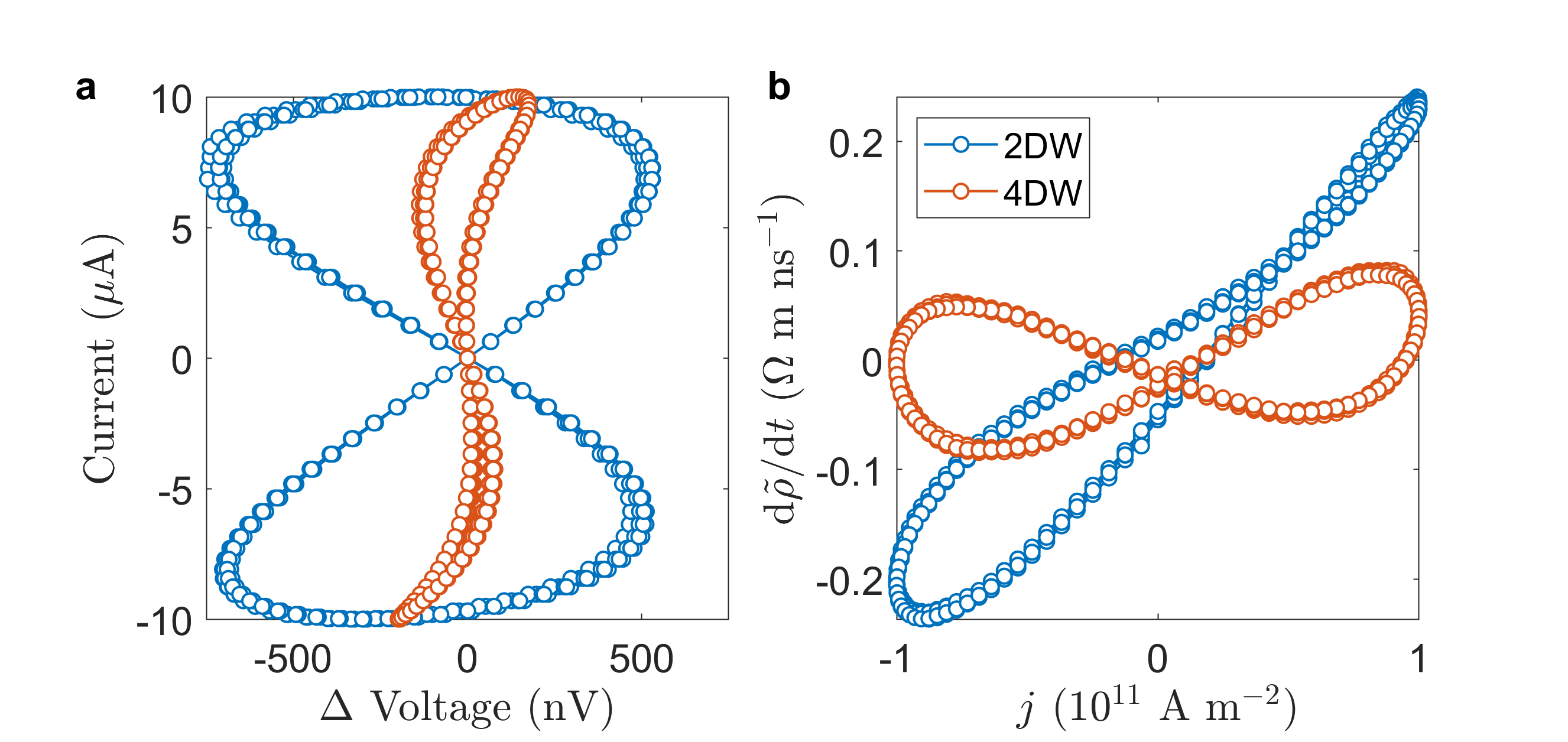}
    \caption{\textbf{Hysteresis curves for domain walls (DW).} Resistivity of the {2DW and 4DW states}. \textbf{(a)} Pinched hysteresis obtained from plotting the device current versus the change in voltage $\Delta V$, in nanovolts. As we can see, the {2DW} state has a larger change in voltage, which is of the order of $500$ nV.
    \textbf{(b)} We plot the change in resistivity as a function of the current density, which is in units of {$10^{11}$~A~m$^{-2}$}. As we can see, the {domain-wall states have}  a larger memristive effect that the ground state and the in-plane states. }
    \label{fig:sims2}
\end{figure}

%This result is consistent with what we derived for the toy model, as $\beta$ depends on the internal magnetization, which is driven by the Zhang-Li coupling as we saw. Since the magnetic superfluid state does not lead to a memristance, we consider a mixed state of superfluid state (MSS), which leads to an AMR effect, and an out of plane components from a dipole, that leads to a memristance and thus an AMM effect. 
While for the cases of magnetic in-plane, out-of-plane and ground states we observe a negligible memristive state, we argue that a non-trivial and measurable effect could be seen for topological magnetic states. In particular, we focus on the  {domain wall states}. %In numerical experiments, these are  are obtained by relaxing hard domain walls, obtaining magnetic states which are equivalent to one and two magnetic kinks, which are however imperfect and thus can have a non-zero memristance in principle. We argue below that the magnetic dipoles interaction is indeed what drives a non-negligible memristive effect, due to the fact that it leads to imperfect superfluid states.
The results for two {such states with two domain walls (2DW) and 4 domain walls (4DW),} and utilizing a current density of $10^{11}$~A/m$^2$ are shown in Figure \ref{fig:sims2}. %, for the superfluid states $n=1$ and $n=2$.
In particular, in Figure \ref{fig:sims2} (a) we plot the current-$\Delta$ Voltage diagram, where this time a non-negligible hysteresis can be observed at the {nano}-volts scale. %The $n=1$ is obtained by relaxing a single hard domain wall, while $n=2$ by relaxing two domain walls. These are then ``imperfect" spin superfluids (which in the toy model would have have zero-memristance) which are dynamically active.
For the {2DW} case, a hysteresis is clearly visible. In Figure \ref{fig:sims2} (b) we plot $\frac{d\tilde \rho}{dt}$ as a function of the current.
We can see that we have a memristance effect with a time varying $\beta$ constant in both the cases {2DW and 4DW}. However, while the curve for the {2DW} is pinched, we have a non-pinched hysteresis for the {4DW} state, which implies a non-zero (and positive) decay  $\eta$. This suggests that we can in principle engineer the behavior of the memristor device depending on the  {annulus}  internal magnetic state.

\begin{figure}
    \centering
    \includegraphics[scale=.5]{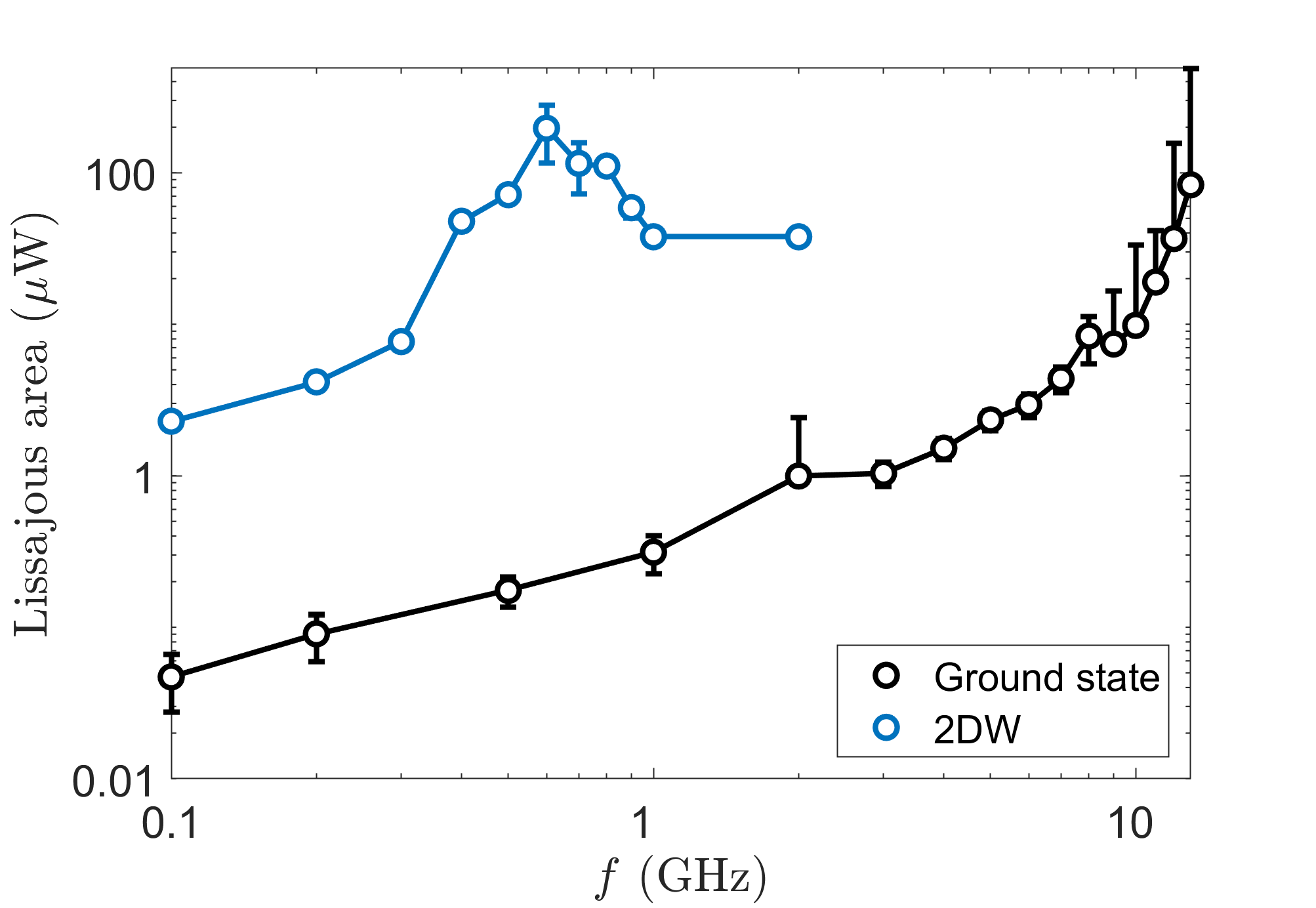}
    \caption{\textbf{Size of the memristive effects for domain walls (DW)}. Area spanned by the curves of the ground state and the {2DW state} in Figs. \ref{fig:sims} and \ref{fig:sims2}. The area provide an estimate of how large is the change of the resistivity  as a function of the frequency of the voltage input per unit of ampere, as in eqn. (\ref{eq:beta}). The peak in Lissajous area for the {2DW state} is at $f=600$~{M}Hz. {The error bars denote standard deviation of the area based on numerical fluctuations and transient effects in the Lissajous' curves.}}
    \label{fig:area}
\end{figure}

\begin{figure}
    \centering
    \includegraphics[scale=.5]{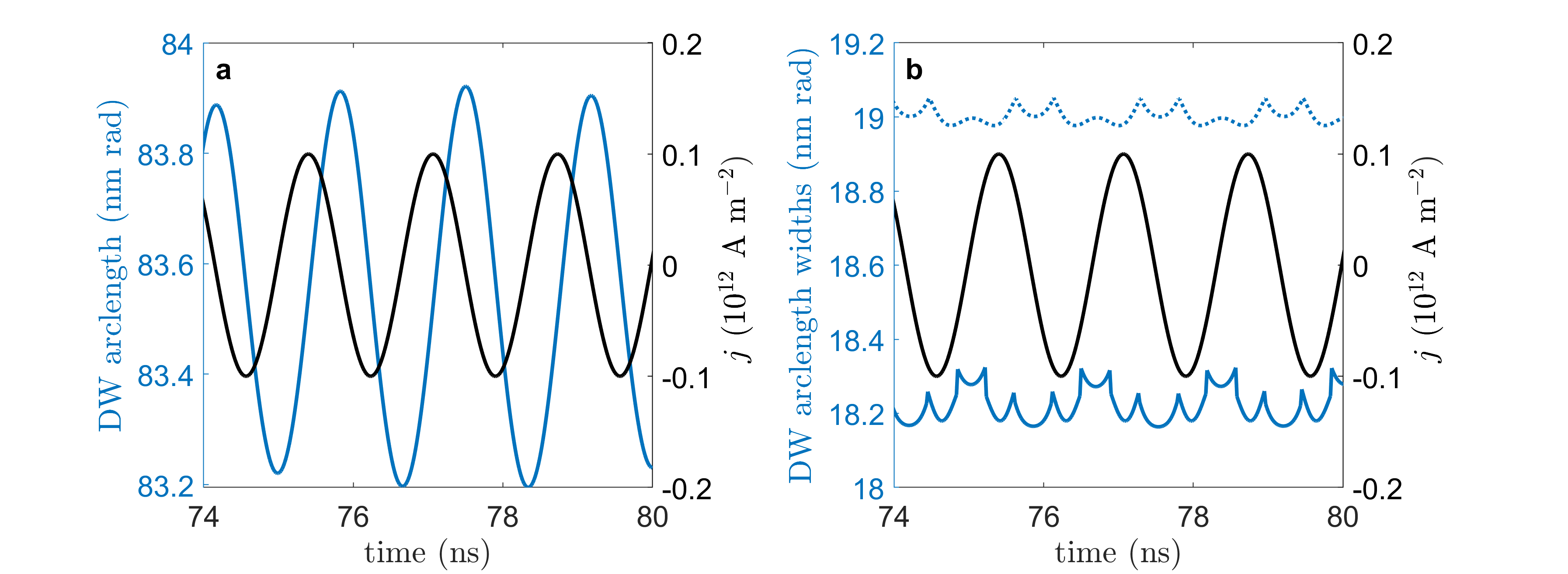}
    \caption{\textbf{Averaged magnetization for domain walls (DW)}.  {Spatially averaged domain-wall dynamics along the path $T(\theta)$ from micromagnetic simulations of the 2DW state subject to a microwave current density $j$ of 1~GHz and amplitude $10\times 10^{11}$~A~m$^{-2}$. The distance between the domain walls is shown to oscillate at the same frequency of the microwave current (black curve) in Panel \textbf{a}. The profile of the domain wall is also modified but exhibiting a more complicated pattern in Panel \textbf{b}. The solid and dotted lines represent each domain wall width of the 2DW state.}}
    \label{fig:path2}
\end{figure}

A way to estimate and compare  how much dependence on the magnetic state we have in the four magnetic states above, is by analysing the  area spanned by the Lissajous curves in $\frac{d\rho}{dt}$ vs. $j$. In Figure \ref{fig:area} we plot the estimated area for curves of the ground state and the {2DW state} of Figure \ref{fig:sims} and \ref{fig:sims2} respectively, which are those with the largest hysteresis in the two sets, as a function of the frequency of the input current density. %As we can see, while the ground state grows with the frequency, the MSS state peaks at around 600 Mhz, which is not far from the Ghz regime.
In both cases, the Lissajous' area for the ground state grows with frequency, a further indication that the memristance in this case is produced by current-induced dynamics. We do not observe a peak up until $10$~GHz.
In contrast, the {2DW} state peaks at around 600~MHz and exhibits a finite magnitude at 1~GHz, which is larger than the magnetic ground state. {The peak at 600~MHz is determined by the most favorable resonance of the coupled 2DW state, determined by the particularities of the energetics. An analytical calculation is beyond the scope of this paper as it necessarily includes the non-local dipole field. However, it is expected that the frequency with maximal Lissajous area will be below ferromagnetic resonance to maintain the spatial localization of domain walls~\cite{Kosevich1990}. If this were not the case, then the domain walls would radiate waves and ultimately annihilate.} It is worth stressing that while the current densities used for the {2DW} state simulations are one order of magnitude smaller than those used for the ground state, the Lissajous area at 600~MHz is approximately two orders of magnitude larger. This demonstrates that the memristance of the {2DW} state is primarily due to the magnetization{'s texture} and in good agreement with a memristive device. The key reason why the memristive effect is amplified is that the magnetic kink cannot be easily removed at small currents, which is due to a form of topological protection. In fact, as the current flows, the kink moves due to the Zhang-Li torque, but cannot be easily destroyed. {Instead, the domain-wall motion induces a periodic breathing dynamic accompanied by slight modifications in the domain-wall profile. These are shown in Fig.~\ref{fig:path2}a-b, respectively, along the path $T(\theta)$. The domain-wall with is estimated by the slope, $m$, of the profile at its center, $2R/m$, where $R$ is the mean radius of the ring. The microwave current is shown in both cases with a solid black curve. We note that the changes are small but the impact on the voltage is larger due to the details of the configuration in 3D.}

\begin{figure}
    \centering
    \includegraphics[scale=.5]{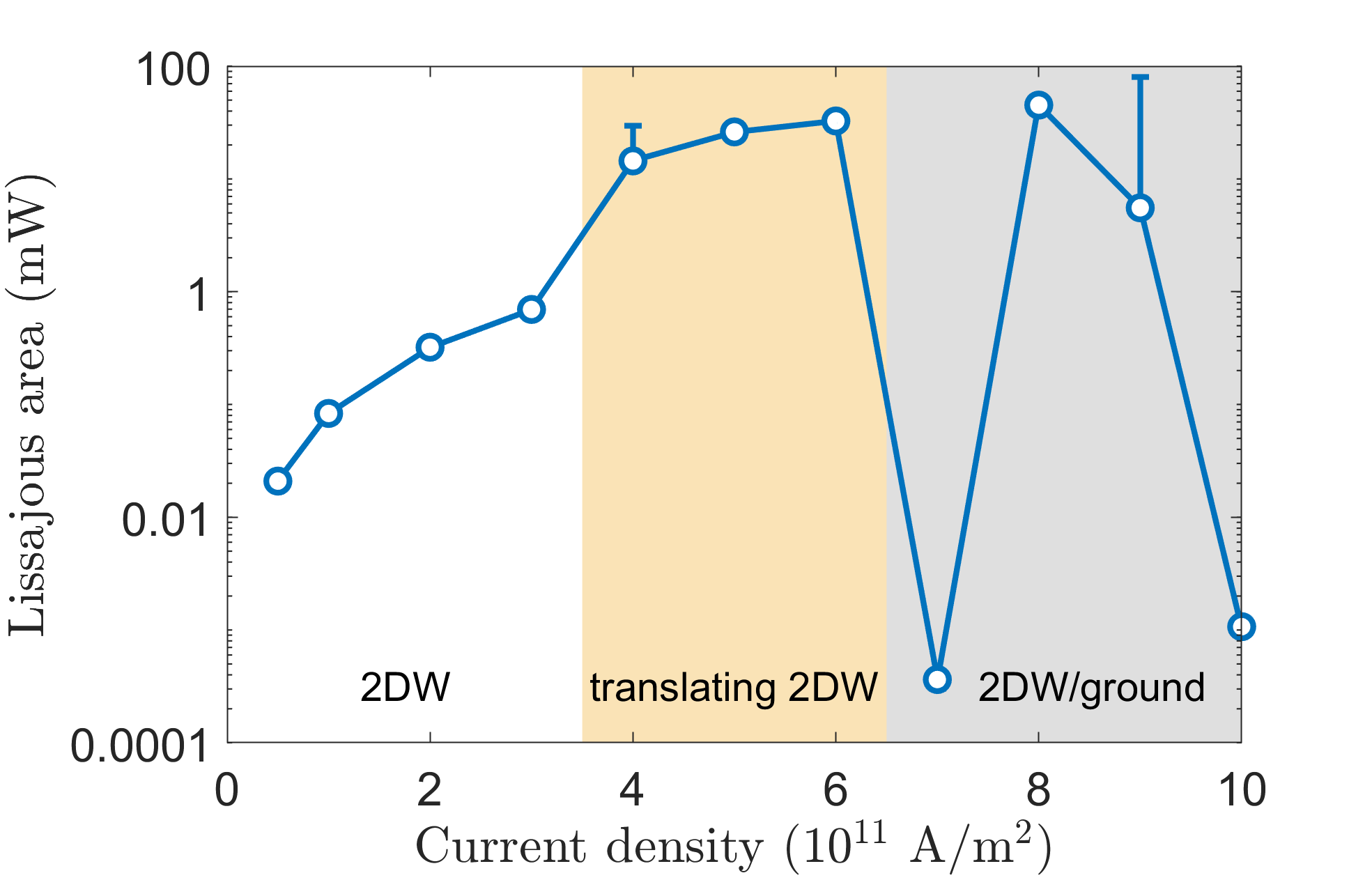}
    \caption{\textbf{Fragility of the 2 domain walls (2DW) state.} We plot the area of the Lissajous curves as a function of the current density. As we can see, the area of the Lissajous curve depends strongly on the current density, and increases as a function of it. At low densities, the {two domain walls are mostly static and} the area is small, indicating a negligible memristive effect. At higher densities {(yellow area), the domain-walls experience a noticeable distortion accompanied by a small yet finite periodic translation of the domain walls. At much higher current densities (gray area), translation dominates and phase slips can occur, leading to the destruction of the domain-wall pair such that the state of the system stabilizes into its ground state. These cases are evidenced by the low Lissajous areas. The stochastic nature of this regime is due to the cell's large aspect ratio. The error bars denote standard deviation of the area based on numerical fluctuations and transient effects in the Lissajous' curves.}}
    \label{fig:frag}
\end{figure}

Based on the {2DW} state, we now explore the current-density dependence on the Lissajous area. The results are presented in Fig.~\ref{fig:frag}. We observe an exponential increase in the area for low current densities that appears to start saturating at $10^{12}$~A~m$^{-2}$. The sudden jump at large current densities is indicative of a sufficient distortion of the {domain walls} to untwist the magnetization, i.e., the topological protection is lost and the system relaxes into the ground state. {The results of this section provide the intuition that at sufficiently low currents, the magnetic states with a finite rotations about the ring's easy axis %non-trivial winding numbers 
are those for which we observe a larger memristive effect. We argue that this observation is associated with the distinct topology of such states when compared to the ground state in the ring. %This is due to the fact that states with a non-trivial winding number are harder to unwrap smoothly and relax to ground states with a negligible memristive behavior. 
As such, this suggests that in real materials topological protection of the magnetic states is an important ingredient for the observation of a dynamical resistance rooted in the AMR effect. The origin of this protection is indeed due to the fact that we are considering a  {annulus}  where these kinks are allowed and not easily annihilated via singularities or phase slips.}

A few {minor} comments are in order.
In addition to the analysis above, we have also tested the dependence on the memristance on the  {annulus}  radius and the thickness. Our numerical studies (see Supp. Mat.) confirms that the thicker the ring, the larger the memristive effect, measured in terms of Lissajous area. Moreover, we have also confirmed that for smaller radii the effect is amplified, in line with the toy model result. However, {It is worth mentioning that the Lissajous areas were only slightly affected for larger  {annulus}  radii. }%Lissajous curves were only slightly affected.
While more in depth studies are required, this is encouraging for experiments relying on nanolithography.

\section*{Discussion}

We have analyzed the memristive effects which occur in realistic magnetic materials due to the interplay between the anisotropic magneto-resistive effect, an electric current via the Zhang-Li coupling, and the magnetodynamics of the material. Our theoretical and numerical results suggest the existence of a memristive effect {in trivial ferromagnets by means of the coupled dynamics of domain-walls. This effect is present when the domain-walls exhibit a non-trivial topology that protects the domain walls from annihilating due to current-induced motion. Our study focused on a  {annulus}  geometry but the presented results are general as to the physical origin of the effect: the interplay between current-induced dynamics and topology. For the purposes of a rigorous derivation and numerical validation, we have specifically }%, of which we have characterized the behavior. Our study focused on a  {annulus}  geometry, and we have obtained a good agreement between theory and numerical simulations. Specifically, we have 
introduced a simple one-dimensional model of a magnetic  {annulus}  with an applied voltage and a resulting current, in which the memristive effect due to the Zhang-Li spin-torque interaction between the current and the local magnetization has been derived analytically. This model characterizes explicitly the type of memrisive behavior the device has, and how the non-adiabatic and adiabatic couplings affect the memristive dynamics, e.g. how the resistance changes with the magnetic state. 

Our results are based on Landau-Lifshitz-Gilbert evolution for the magnetization applied to the Anisotropic-Magnetoresistance effect in continuous magnetic media.
We focused on the study of a common magnetic material such as permalloy, which is readily available. The model explicitly shows that a memristive effect should be present in soft-magnetic material. We  predict that for permalloy, it should be measurable at a nano-Volt scale and in the range of $600-1000$~MHz. 

One of the most striking features of this device is that the behavior of the memristance can be engineered by inducing particular magnetic states.  While the effect can be considered small by the standard of typical oxide-devices, we argue that a larger effect should be present in {annulii}-based lattices, %{\textbf{Ezio: why heterostructures? Do you mean instead lattices?}}, 
as the change in voltage should accumulate in a coherent fashion. From an experimental perspective, our results are valid for AC currents, implying that for such effect to be observable any DC voltage should be filtered. However, this should be possible using a setup with capacitors or lock-in amplifiers, also given the fact that the effect is in the nV regime. The main experimental challenge might be to stabilize a full rotation of the magnetization in a device that necessarily will depart from a perfect ring, {such as a horseshoe geometry,} in order to manipulate the magnetization and to inject a microwave current. %{\textbf{\textcolor{green}{GIA-WEI}, can you comment here? Or do you agree? Not sure if a ``horseshoe'' type of device could host this.}}

From the perspective of the memristive device, engineering the hysteresis is an exciting development. Typically, the hysteresis curves of memristive devices strongly depend on the type of underlying physical properties of the device \cite{stru13}. While even in numerical experiments we can see that our device is not ideal, we have observed that depending on the magnetic states we can have a non-current dependent memristance decay which is zero. 

Crucially, the underlying feature of the amplified memristive effect is a topological protection, {enacted by the similar chirality of the domain walls that are only removable by the formation of singularities.} %due to the non-removability of the kink in the mixed magnetic superfluid $n=1$.
The proposed memristor is therefore topological in nature. The possibility of \textit{a posteriori} re-programming the memristor by changing the magnetic properties of the device makes this type of device  ideal for a fast neuromorphic Random Access Memory.

This proposal is a departure from previous works on the subject.
In fact, other memristive devices based on magnetic states used  viscosity to obtain a memristive state, for instance via  magnetic pinning \cite{allwood}, domain-wall motion \cite{cargwcnis}, inclusion of magnetic tunnel junctions \cite{grollier1}, which induce a slower dynamics.
Instead, here we have shown that at the natural timescale (GHz) typical of magnetic dynamics there can be a memristive effect insofar as the stabilized magnetic state is topological. %Thus, our paper provides a direct connection between topology and electrical memory effects in magnetic devices, providing an analytical foundation.

The present study opens new analyses in this direction. We have analyzed only a handful of magnetic states of a ring. %In  {annulus}  heterostructures many more possible and non-trivial interactions between the magnetic states are possible, and include the formation and propagation of spin waves. 
{In  {annulus}  heterostructures composed of arrays of interconnected rings, the dynamics may interfere additively to increase the total change in the AMR and thus the size of the AMM effect. In other words, we expect that larger structures composed of nanorings will amplify such an effect. Moreover, because we observe an essential topological effect, we expect that such a phenomenon might be observable also in large lattice structures. For instance, by joining rings in a 1d structure, we expect that for large enough currents domain walls can be formed at the junctions. Then, according to our paper, this should correspond to a memristive effect with a threshold dynamics, e.g. for low enough voltages there is no memristance, but at a certain (material dependent) voltage such hysteresis should be expected. This is also the general conclusion of \cite{cargwcnis}, however for a rather simplified toy model.%Moreover, we have considered the case of permalloy, but different and harder magnetic materials may  behave differently and provide a larger memristive effect.  %All these features will be considered in future studies.
Another possibility is to introduce or combine other magnetoresistive effects. For example, GMR would provide a much larger signal but it will necessarily require coupled magnetic layers. In this case, the coupling between solitons must be carefully studied and related to GMR from the point of view of memristance. It would interesting to clarify whether a memristive effect would exist at GHz frequencies in this case.}

{Overall, while this paper does not provide an exhaustive list of magnetic states and geometries that maximize such complex interplay of phenomena, it provides a foundation for the study of this effect. As we have discussed, the memristance effect arises from the combination of dynamical magnetic and non-trivial magnetic states which cannot be easily annihilated due to their topological nature. Based on our analysis, we expect other geometries to satisfy both conditions. A situation of interest is found in 360 degrees domain walls in nanowires. This geometry would provide both a much simpler means to experimentally set a current flow due to its two terminals and an excitation mechanism to produce domain walls at will. Indeed, this geometry has been theoretically studied to date in the context of spin superfluidity~\cite{Iacocca2017,Sonin2010,Takei2014,Iacocca2019b}, which can be interpreted as a train of homochiral domain-walls. When two domain walls are present, this state is the 1D projection of the 2DW state studied here. Therefore, the same anisotropic magnetomemristance effect is expected. In addition, initial evidence of spin superfluids in antiferromagnets~\cite{Yuan2018, Stepanov2018} also suggests that these effect could be generalized to other materials. Finally, it would be also interesting to explore a two-dimensional version of this effect in lattices of topological objects, such as skyrmion lattices~\cite{Muhlbauer2009,Montoya2017}. A similar rationale of topology and dynamics suggests that these lattices as a whole could exhibit memristance.}

%Overall, while this paper does not provide an exhaustive answer of which magnetic states maximize such complex interplay of phenomena, it provides the foundation for the study of this effect. As we have discussed, the memristance effect arises from the combination of dynamical magnetic and non-trivial magnetic states which cannot be easily annihilated due to their topological nature. It would be interesting to explore if other topological states exhibit memristance, for example, spin superfluids or dissipative exchange flows in nanowires~\cite{Sonin2010,Takei2014,Iacocca2017,Iacocca2019b, Yuan2018, Stepanov2018} or even skyrmion lattices in extended films~\cite{Muhlbauer2009,Montoya2017}. %This paper shows that it is in fact not enough to have simple domain wall distortion, but one needs non-trivial magnetic states.}

%\textbf{Ezio: a thought is to discuss the many many different numerical simulations that can be done. What happens when MSS' n increases, what happens with other materials, stacking the disks, etc.? Also, from an experimental point of view, it would be good to point out that one would need a high-pass filter, or at least a tiny capacitor, to filter out the DC. Alternatively, one could use a lock-in amplifier to measure only the 2f component or even an oscilloscope to take time-traces. The frequencies are well within the measurable capabilities I believe. As for nV, this is certainly possible in DC, so I would guess one can use lock-in amplifier + nanovoltmeter combo to achieve this.}

\section*{Methods}

\subsection*{Sketch of the theoretical derivations}
We provide a sketch of the proof of the main formulae provided in the paper, in particular eqns. (\ref{eq:memlike})-(\ref{eq:aux}). {The full derivations are provided in Supplementary Note 1 of the Supplementary Material.}
The key equation which we use as a starting point is the Landau-Lifschitz-Gilbert equation:

\begin{eqnarray}
%\frac{\partial\mathbf{M}}{\partial t}&=&
%-\gamma\mathbf{M}\times\vec{\mathbf{H}}_\text{eff}
%+\frac{\frac{\alpha}{M_s}}{M_\text{S}}\mathbf{M}\times\frac{\partial\mathbf{M}}{\partial t} \nonumber \\
%&+&\frac{pa^3}{2eM_\text{S}}\mathbf{M}\times\big((\mathbf{J}(\mathbf{r})\cdot\nabla)\mathbf{M}\big),
\frac{\partial\hat{\mathbf{m}}}{\partial t}&=&
-\gamma\mu_0\hat{\mathbf{m}}\times\vec{\mathbf{H}}_\text{eff}
+\alpha\hat{\mathbf{m}}\times\frac{\partial\hat{\mathbf{m}}}{\partial t} \nonumber \\
&-&\xi \nu\left(\hat{\mathbf{m}}\times(\vec{\mathbf{J}}(t)\cdot\nabla)\hat{\mathbf{m}}\right)\nonumber \\&-&\nu\hat{\mathbf{m}}\times\left(\hat{\mathbf{m}}\times(\vec{\mathbf{J}}(t)\cdot\nabla)\hat{\mathbf{m}}\right)
\label{eq:LLG2}
\end{eqnarray}
and the AMR correction to the electric field
\begin{equation}
\vec{\mathbf{E}}=\rho_0 \vec{\mathbf{J}}+\hat {\mathbf{m}} ( \rho_{||} -\rho_{\perp}) (\hat{ \mathbf{m}}\cdot \vec{\mathbf{J}}).
\label{eq:amr2}
\end{equation}
The voltage drop along a certain path is given by
\begin{eqnarray}
V(t)&=&\int_{\gamma} \vec{\mathbf{E}}(\vec{\mathbf{J}}(t))\cdot\hat{\mathbf{T}}(\gamma) d\gamma.
\label{eq:efield}
\end{eqnarray}
Substituting eqn. (\ref{eq:amr2}) in eqn. (\ref{eq:efield}) one obtains the voltage in terms of the magnetization.
\begin{eqnarray}
V(t)&=&\int_{\gamma} \big(\rho_0 \vec{\mathbf{J}}+\hat {\mathbf{m}} ( \rho_{||} -\rho_{\perp}) (\hat{ \mathbf{m}}\cdot \vec{\mathbf{J}})\big)\cdot\hat{\mathbf{T}}(\gamma).
\label{eq:efield2}
\end{eqnarray}
If the current is along the  {annulus}  then as discussed in the main text, we have
\begin{eqnarray}
\vec {\mathbf{J}}(t)=j(t)\hat{\mathbf{T}}(\theta).
\end{eqnarray}
Thus, we can define the resistance along a ring, we can define the path $\gamma$ as $d\gamma=r d\theta$, obtaining
\begin{eqnarray}
R(t)=\frac{V(t)}{j(t)}&=&2\pi r \rho_0+r\Delta \rho\int d\theta \big(  \hat{ \mathbf{m}}\cdot \hat{\mathbf{T}}(\theta)\big)^2.
\label{eq:efield2}
\end{eqnarray}
which is the formula we provided in the main text. Eqns. (\ref{eq:memlike})-(\ref{eq:aux}) can now be derived taking a time derivative, and replacing $\frac{\partial\hat{\mathbf{m}}}{\partial t}$ with the LLG equation.
\subsection*{Numerical Simulations}

Micromagnetic simulations using MuMax3~\cite{Vansteenkiste2014} were performed on a GPU NVIDIA Geforce 940MX. The system is discretized in cells of approximately 1~nm~$\times$1~nm~$\times$10~nm. While the aspect ratio is large, it has been shown in previous works that the numerical effects are negligible for near-uniform states across the thickness~\cite{Iacocca2020}. We note that possible instabilities across the thickness would be hidden, but we do not expect such instabilities to exist in uniform or topological states insofar as the magnetization is largely in plane or relatively large in extent~\cite{Iacocca2017}. For large current densities, we do observe the compression of the domain wall to an extent where the magnetization is out-of-plane and a phase-slips occur. Such a phase-slip could occur at smaller currents for cubic cells, i.e., via a texture distortion across the thickness. %\textbf{If this smells bad, I can run a thinner film for comparison. It will only scale the dipole... Unfortunately, I am tied to 2D simulations until my workstation arrives or the local supercomputer is built}. 
The simulation is set to use a Runge-Kutta 45 stepper and the maximum step time is set to 100$^\mathrm{th}$ of a period. This is essential to avoid noise for simulations excited by currents of different frequencies.

It is fundamental to stabilize a robust initial condition. The general protocol followed features a ``tiered'' energy minimization approach to quench spurious dynamics down to numerical accuracy: we first use MuMax's built-in energy minimization function \texttt{relax}, then we run for $20$~ns with $\alpha = 1$, $30$~ns with $\alpha = 0.1$ and $50$~ns with $\alpha = 0.01$. The current density is implemented as being parallel to $\mathbf{T}(\theta)$. Such current density is generated in Matlab with and then imported to MuMax3 as a mask whose magnitude is time-dependent.

%Dynamic simulations are run for 20 to 100 periods, depending on the frequency. In general, larger frequencies require more time to stabilize as sudden introduction of current launches spin waves from the texture that need to attenuate to discern a memristive effect. The resulting time-dependent magnetization
{Dynamic simulations are run for 20 to 100 periods, depending on the frequency of the current. In general, larger frequencies require more time to stabilize into periodic dynamics. This is a consequence of the sudden introduction of current launches spurious spin waves that interact and move the domain wall without resistance because of the lack of material imperfections in the simulations. The resulting time-dependent magnetization dynamics }is then processed in MATLAB to obtain the resistivity according to Eqn.~\eqref{eq:res}. The change in resistivity $d\tilde{\rho}/dt$ is calculated as the numerical derivative of Eqn.~\eqref{eq:res} while the voltage is simply extracted from Eqn.~\eqref{eq:res} and we consider the average radius of the ring.

{The dynamics along the path $T(\theta)$ are extracted from the micromagnetic simulations by transforming space from Cartesian into polar coordinates. A circumference is identified from the midpoint of the ring, within one micromagnetic cell. The resulting data is composed of the in-plane magnetization components as a function of the angle $\theta$. Finally, we implement a rotation matrix to obtain the in-plane magnetization components in polar coordinates, which we identify as $m_t$ (tangential) and $m_p$ (perpendicular).}

\subsection*{Data availability}

The data that support the findings of this study are available from the corresponding
author upon reasonable request

\printbibliography %Prints bibliography

\section*{Acknowledgements.} The work of FC and CN was carried out under the auspices of the US Department of Energy through the Los Alamos National Laboratory, operated by Triad National SecurityLLC (Contract No. 892333218NCA000001).
%\section*{Funding } 
CN was funded by DOE-LDRD grant 2017014ER. FC was also financed via DOE-LDRD grants PRD20170660 and PRD20190195. CILA was funded by CNPq 432029/2018-1, FAPEMIG, Coordena\c{c}\~{a}o de Aperfei\c{c}oamento de Pessoal de N\'{i}vel Superior (CAPES) - Finance Code 001 and Newton Fund NAF-R2-192040.

\section*{Competing interests}
The authors declare no competing interests.

\section*{Author contributions} 
C.I.L.A and E.I performed the numerical simulations and F.C developed and worked on the theoretical model.  F.C., C. N., C.I.L.A. E. I. and G.W.C. contributed to the discussions on analytical and numerical results, as well as writing the paper. F.C. and E. I. contributed equally to this work.%the numerical experiments, the models to explain it and writing the paper. 

\appendix
\section*{Supplementary Note 1}
%\subsection{Spin-Transfer torque memristor: Ring model}
Let us consider the following simplified model. We consider a  {ring} of length $L=2\pi r$, where $r$ is the radius, {but of negligible thickness}. We consider a magnetization vector dependent on the angle on the ring, $\vec{\boldsymbol{M}}_t(\theta)=M_x(t,\theta)\boldsymbol{\hat x}+M_y(t,\theta) \boldsymbol{\hat y}+M_z(t,\theta)\boldsymbol{ \hat z}$. The resistivity of the alloy per unit of length is $\rho$, such that $R=2\pi \rho  r$. We also consider a voltage generator $V_0$ on the ring, which introduces a current $j$ in the ring, such that $V_0=Rj$. Vectorially, we have that the vector of the radius is given by
\begin{eqnarray}
\boldsymbol{r}=r [\cos\theta, \sin \theta].
\end{eqnarray}
Since $j$ flows on the ring, we must have $\boldsymbol{r}\cdot \vec{\boldsymbol{J}}=0$, which implies $ \vec{\boldsymbol{J}}(t,\theta)=j(t) [-\sin \theta,\cos \theta,0]=j\hat{\boldsymbol{T}}(\theta)$, asssuming that the current is flowing clockwise if $j>0$, and where $\hat{\boldsymbol{T}}(\theta)$ is the tangent to the wire, with the property $\hat{\boldsymbol{T}}(\theta)\cdot \hat{\boldsymbol{T}}(\theta)=1$. 
The equation for the magnetization is given by the LLG equation \cite{Gilbert} plus the Zhang-Li coupling with the current \cite{Zhang2004}:
\begin{eqnarray}
\frac{\partial\vec{\boldsymbol{M}}_t}{\partial t}&=&
-\gamma\vec{\boldsymbol{M}}_t\times\vec{\boldsymbol{H}}_\text{eff}
+\frac{\alpha}{M_s}\vec{\boldsymbol{M}}_t\times\frac{\partial\vec{\boldsymbol{M}}_t}{\partial t}\nonumber \\ &-&\frac{\nu}{M_S}\Big(\xi\vec{\boldsymbol{M}}_t\times +\frac{1}{M_S} \vec{\boldsymbol{M}}_t\times\vec{\boldsymbol{M}}_t\times\Big) \Big(\big(\vec{\boldsymbol{J}}(\boldsymbol{\theta})\cdot\triangledown\big)\vec{\boldsymbol{M}}_t\Big),
\end{eqnarray}
which we need to reduce on an effective one dimensional model by converting the $\nabla$ operator on the ring. 
We have $\vec{\boldsymbol{J}}(\boldsymbol{\theta})\cdot\triangledown=j_x \partial_x+j_y \partial_y$, and we can write $\partial_{\cdot} \vec{\boldsymbol{M}}_t(\theta)=(\partial_{\cdot} \theta) \partial_\theta \vec{\boldsymbol{M}}_t(\theta)$. 

We have $\theta=\arctan \frac{y}{x}$, from which we obtain $\partial_x \theta=-\frac{y}{x^2+y^2}=-\frac{1}{r} \sin \theta$ and $\partial_y=\frac{x}{x^2+y^2}=\frac{1}{r} \cos\theta$. It follows that $(j_x\partial_x+j_y \partial_y)\vec{\boldsymbol{M}}_t(\theta)=\frac{j}{r}(\cos\theta^2+\sin\theta^2) \partial_\theta \vec{\boldsymbol{M}}_t(\theta)=\frac{j}{r} \partial_\theta \vec{\boldsymbol{M}}_t(\theta)$. We thus obtain that, on the ring, we have the reduced equation
\begin{eqnarray}
\frac{\partial\vec{\boldsymbol{M}}_t(\theta)}{\partial t}%&=&
%-\gamma\vec{\boldsymbol{M}}_t(\theta)\times\vec{\boldsymbol{H}}_\text{eff}
%+\frac{\frac{\alpha}{M_s}}{M_\text{S}}\vec{\boldsymbol{M}}_t(\theta)\times\frac{\partial\vec{\boldsymbol{M}}_t(\theta)}{\partial t} +\frac{pa^3 j }{2eM_\text{S} r}\vec{\boldsymbol{M}}_t(\theta)\times\frac{\partial \vec{\boldsymbol{M}}_t(\theta)}{\partial \theta} \nonumber \\
=\vec{\boldsymbol{M}}_t(\theta)\times \Big( &-&\gamma \vec{\boldsymbol{H}}_\text{eff}+\frac{\alpha}{M_s} \frac{\partial\vec{\boldsymbol{M}}_t(\theta)}{\partial t} -\frac{\xi \nu j}{M_\text{S} r}\frac{\partial \vec{\boldsymbol{M}}_t(\theta)}{\partial \theta}\nonumber \\
&-&\frac{\nu j}{M_S^2}\vec{\boldsymbol{M}}_t(\theta)\times \frac{\partial \vec{\boldsymbol{M}}_t(\theta)}{\partial \theta} \Big),
\label{eq:llgring}
\end{eqnarray}
where we neglect all finite-size effects, and in the last line we used the notation employed by MuMax.

The AMR effect is given by the vector field in the wire
\begin{eqnarray}
\vec{\boldsymbol{E}}=\rho_0 \vec{\boldsymbol{J}}+\Delta \rho (\vec{\boldsymbol{J}}\cdot \hat{\boldsymbol{m}}_t(\theta)) \hat{\boldsymbol{m}}_t(\theta).
\end{eqnarray}
The effective voltage is given by
\begin{eqnarray}
V=r \int_0^{2\pi} \vec{\boldsymbol{E}}\cdot \hat{\boldsymbol{T}}(\theta)\ d\theta.
\end{eqnarray}
If we replace now the AMR correction, we find
\begin{eqnarray}
V&=&r \int_0^{2\pi} \big(\rho_0 \vec{\boldsymbol{J}}+\Delta \rho (\vec{\boldsymbol{J}}\cdot \hat{\boldsymbol{m}}_t(\theta))\hat{\boldsymbol{m}}_t(\theta)\big)\cdot \hat{\boldsymbol{T}}(\theta)\ d\theta  \nonumber \\
&=&r \int_0^{2\pi} \big(\rho_0 j \hat{\boldsymbol{T}}(\theta)+j\Delta \rho  (\hat{\boldsymbol{T}}(\theta)\cdot \hat{\boldsymbol{m}}_t(\theta))\hat{\boldsymbol{m}}_t(\theta)\big)\cdot \hat{\boldsymbol{T}}(\theta)\ d\theta  \nonumber \\
&=&r \int_0^{2\pi} \big(\rho_0 j+j\Delta \rho  (\hat{\boldsymbol{T}}(\theta)\cdot \hat{\boldsymbol{m}}_t(\theta))^2\big)\ d\theta \nonumber \\
&=&V_0+j r(\Delta \rho)\int_0^{2\pi}  (\hat{\boldsymbol{T}}(\theta)\cdot \hat{\boldsymbol{m}}_t(\theta))^2\ d\theta
\end{eqnarray}
where we used  $\hat{\boldsymbol{T}}(\theta)\cdot \hat{\boldsymbol{T}}(\theta)=1$.
We can write explicitly
\begin{eqnarray}
\hat{\boldsymbol{T}}(\theta)\cdot \vec{\boldsymbol{M}}_t(t,\theta)=M_y(t,\theta) \cos\theta-M_x(t,\theta) \sin(\theta),
\label{eq:extra}
\end{eqnarray}
where $M_y^2+M_x^2=M_S^2$.
Let us now note that
\begin{eqnarray}
V=j\Big(\rho_0 2\pi r+r\Delta \rho\int_0^{2\pi}  (\hat{\boldsymbol{T}}(\theta)\cdot \hat{\boldsymbol{m}}_t(\theta))^2\ d\theta\Big)=j \tilde R,
\end{eqnarray}
where $\tilde R$ is an effective resistance. If $\vec{\boldsymbol{M}}_t(\theta)$ does not depend on $j$, then $\tilde R$ is only a constant with respect to the history of the voltage, and thus is simply a resistance. If on the other hand in eqn. (\ref{eq:llgring}) we have $p\neq 0$, clearly there is a history dependence in $M$ on $j$, and the system has a memory.

\subsection*{Supplementary Note 2} 

Let us now look at the effective resistance, which is
\begin{eqnarray}
\tilde R=\frac{V(t)}{j(t)}=R_0+ \frac{r \Delta \rho}{M_S^2} \int_0^{2\pi}  (\hat{\boldsymbol{T}}(\theta)\cdot \vec{\boldsymbol{M}}_t(\theta))^2\ d\theta
\end{eqnarray}
If we take a derivative with respect to time, we obtain, and define $ \Delta\vec{\boldsymbol{M}}=-\gamma \vec{\boldsymbol{H}}_\text{eff}+\frac{\alpha}{M_s} \frac{\partial\vec{\boldsymbol{M}}_t(\theta)}{\partial t} -\frac{\xi \nu }{M_\text{S} r}\frac{\partial \vec{\boldsymbol{M}}_t(\theta)}{\partial \theta}-\frac{\nu}{M_S^2 r} \vec{\boldsymbol{M}}_t(\theta)\times  \frac{\partial \vec{\boldsymbol{M}}_t(\theta)}{\partial \theta}$
\begin{eqnarray}
\partial_t \tilde R&=&\frac{r \Delta \rho}{M_S^2} \int_0^{2\pi} ( \hat{\boldsymbol{T}}(\theta)\cdot  \vec{\boldsymbol{M}}_t(\theta))(\hat{\boldsymbol{T}}(\theta)\cdot \partial_t \vec{\boldsymbol{M}}_t(\theta))\ d\theta\nonumber \\
&=&\frac{r \Delta \rho}{M_S^2} \int_0^{2\pi} ( \hat{\boldsymbol{T}}(\theta)\cdot  \vec{\boldsymbol{M}}_t(\theta))(\hat{\boldsymbol{T}}(\theta)\cdot \Big(\vec{\boldsymbol{M}}_t(\theta)\times \Delta\vec{\boldsymbol{M}}\Big)\Big)\ d\theta\nonumber \\
&=&\frac{r \Delta \rho}{M_S^2} \int_0^{2\pi} ( \hat{\boldsymbol{T}}(\theta)\cdot  \vec{\boldsymbol{M}}_t(\theta))(\Delta\vec{\boldsymbol{M}}\cdot \Big(\hat{\boldsymbol{T}}(\theta)\times \vec{\boldsymbol{M}}_t(\theta) \Big)\Big)\ d\theta.
\end{eqnarray}

From which we see that if the magnetization is either parallel or perpendicular to $\hat{\boldsymbol{T}}$ the resistance does not change. Similarly, if the vector $\Delta \vec{\boldsymbol{M}}$ is perpendicular to either $ \hat{\boldsymbol{T}}$ or $\vec{\boldsymbol{M}}$ along the ring, the resistance does not change. The memristive effect is possible in the regime in which $\|\vec{\boldsymbol{H}}_{eff}\|\ll\frac{\nu }{M_\text{S} r\gamma }\|\frac{\partial \vec{\boldsymbol{M}}_t(\theta)}{\partial \theta}\| $ and the system's dissipation is small. Thus, the effective memristor equation is given by
\begin{eqnarray}
\frac{\partial \tilde R}{\partial t}
&=&-\frac{ \nu j(t) \Delta \rho}{M_\text{S}^3  }  \int_0^{2\pi} ( \hat{\boldsymbol{T}}(\theta)\cdot  \vec{\boldsymbol{M}}_t(\theta))\Big(\big(\xi \frac{\partial \vec{\boldsymbol{M}}_t(\theta)}{\partial \theta}+\frac{1}{M_S} \vec{\boldsymbol{M}}_t(\theta)\times \frac{\partial \vec{\boldsymbol{M}}_t(\theta)}{\partial \theta}\big)\cdot \Big(\hat{\boldsymbol{T}}(\theta)\times \vec{\boldsymbol{M}}_t(\theta) \Big)\Big)\ d\theta \nonumber \\
&-& \frac{r \Delta \rho}{M_S^2} \int_0^{2\pi} ( \hat{\boldsymbol{T}}(\theta)\cdot  \vec{\boldsymbol{M}}_t(\theta))((\gamma \vec{\boldsymbol{H}}_\text{eff}-\frac{\alpha}{M_s} \frac{\partial\vec{\boldsymbol{M}}_t(\theta)}{\partial t})\cdot \big(\hat{\boldsymbol{T}}(\theta)\times \vec{\boldsymbol{M}}_t(\theta) \big)\Big)\ d\theta \nonumber \\
&\equiv& \beta(\vec{\boldsymbol{M}}_t) j(t)-  \eta(\vec{\boldsymbol{M}}_t,\vec{\boldsymbol{H}}_{eff})
\label{eq:memlike_full}
\end{eqnarray}
 where we used the identity $\boldsymbol{a}\cdot (\boldsymbol{b} \times\boldsymbol{c})= \boldsymbol{c}\cdot (\boldsymbol{a} \times\boldsymbol{b})=\boldsymbol{b}\cdot (\boldsymbol{c} \times\boldsymbol{a})$.
 We see that we have an dynamical memristor equation in which the internal parameter is the magnetization vector along the ring. It is interesting to note that the geometry of the ring, which is its radius $r$, enters only in the parameter $\alpha$ and cancels in the parameter $\beta$.
We can now rescale the value of $\vec{\boldsymbol{M}}_t(\theta)=M_S  \hat{\boldsymbol{m}}_t$ (or equivalently setting $M_S=1$). In terms of the normalized magnetic field $\hat{\boldsymbol{m}}(\theta)$ , we have
% \begin{eqnarray}
%     \eta(\hat{\boldsymbol{m}}_t,\vec{\boldsymbol{H}}_{eff})&=& r \Delta \rho \int_0^{2\pi} ( \hat{\boldsymbol{T}}(\theta)\cdot  \hat{\boldsymbol{m}}_t(\theta))((\gamma \vec{\boldsymbol{H}}_\text{eff}-\frac{\alpha}{M_s} \frac{\partial\hat{\vec{\boldsymbol{M}}}_t(\theta)}{\partial t})\cdot \big(\hat{\boldsymbol{T}}(\theta)\times \vec{\boldsymbol{M}}_t(\theta) \big)\Big)\ d\theta\\
%     \beta(\hat{\vec{\boldsymbol{M}}}_t)&=&\frac{ \nu  \Delta \rho}{M_\text{S}  }  \int_0^{2\pi} ( \hat{\boldsymbol{T}}(\theta)\cdot  \vec{\boldsymbol{M}}_t(\theta))\Big(\big(\xi \frac{\partial \hat{\vec{\boldsymbol{M}}}_t(\theta)}{\partial \theta}+\frac{1}{M_S} \vec{\boldsymbol{M}}_t(\theta)\times \frac{\partial \vec{\boldsymbol{M}}_t(\theta)}{\partial \theta}\big)\cdot \Big(\hat{\boldsymbol{T}}(\theta)\times \vec{\boldsymbol{M}}_t(\theta) \Big)\Big)\ d\theta
 %\end{eqnarray}
% Obtaining
 \begin{eqnarray}
     \eta(\hat{\boldsymbol{m}}_t,\vec{\boldsymbol{H}}_{eff})&=& r  \Delta \rho \int_0^{2\pi} ( \hat{\boldsymbol{T}}(\theta)\cdot  \hat{\boldsymbol{m}}_t(\theta)(\theta))((\gamma \vec{\boldsymbol{H}}_\text{eff}-\alpha \frac{\partial\hat{\boldsymbol{m}}_t(\theta)(\theta)}{\partial t})\cdot \big(\hat{\boldsymbol{T}}(\theta)\times \hat{\boldsymbol{m}}_t(\theta)(\theta) \big)\Big)\ d\theta\\
     \beta(\hat{\boldsymbol{m}}_t(\theta))&=&  \nu  \Delta \rho   \int_0^{2\pi} ( \hat{\boldsymbol{T}}(\theta)\cdot  \hat{\boldsymbol{m}}_t(\theta)(\theta))\Big(\big(\xi \frac{\partial \hat{\boldsymbol{m}}_t(\theta)}{\partial \theta}+ \hat{\boldsymbol{m}}_t(\theta)\times \frac{\partial \hat{\boldsymbol{m}}_t(\theta)}{\partial \theta}\big)\cdot \Big(\hat{\boldsymbol{T}}(\theta)\times \hat{\boldsymbol{m}}_t(\theta) \Big)\Big)\ d\theta\nonumber 
 \end{eqnarray}
Let us now use the fact that
\begin{eqnarray}
\big(\hat{\boldsymbol{m}}_t(\theta)\times \frac{\partial \hat{\boldsymbol{m}}_t(\theta)}{\partial \theta}\big)\cdot \big(\hat{\boldsymbol{T}}(\theta)\times \hat{\boldsymbol{m}}_t(\theta) \big)&=&\hat{\boldsymbol{T}}(\theta) \cdot \Big(\hat{\boldsymbol{m}}_t(\theta)\times  \big( \hat{\boldsymbol{m}}_t(\theta) \times \frac{\partial \hat{\boldsymbol{m}}_t(\theta)}{\partial \theta} \big)\Big)
\end{eqnarray}
We now use the identity $\boldsymbol{a}\times (\boldsymbol{b} \times \boldsymbol{c})=(\boldsymbol{a}\cdot \boldsymbol{c})\boldsymbol{b}-(\boldsymbol{a}\cdot \boldsymbol{b}) \boldsymbol{c}$, to obtain
\begin{eqnarray}
\hat{\boldsymbol{T}}(\theta) \cdot \Big(\hat{\boldsymbol{m}}_t(\theta)\times  \big( \hat{\boldsymbol{m}}_t(\theta) \times \frac{\partial \hat{\boldsymbol{m}}_t(\theta)}{\partial \theta} \big)\Big)=(\hat{\boldsymbol{T}}(\theta)\cdot \hat{\boldsymbol{m}}_t)(\frac{\partial \hat{\boldsymbol{m}}_t(\theta)}{\partial \theta} \cdot \hat{\boldsymbol{m}}_t(\theta))-\|\hat{\boldsymbol{m}}_t(\theta)\|^2(\hat{\boldsymbol{T}}(\theta)\cdot \frac{\partial \hat{\boldsymbol{m}}_t(\theta)}{\partial \theta}).
\end{eqnarray}
If we use that
$||\hat{\boldsymbol{m}}_t(\theta)||^2=1$, we define $t_m(\theta)=\hat{\boldsymbol{T}}(\theta)\cdot \hat{\boldsymbol{m}}_t(\theta)$ 
and from which we see that we can rewrite the parameter $\beta$ as the expression
\begin{eqnarray}
\frac{\beta(\hat{\boldsymbol{m}}_t)}{ \nu \Delta \rho }&=&- \xi \  \int_0^{2\pi} t_m(\theta)\Big( \frac{\partial \hat{\boldsymbol{m}}_t(\theta)}{\partial \theta}\cdot \Big(\hat{\boldsymbol{T}}(\theta)\times \hat{\boldsymbol{m}}_t(\theta) \Big)\Big)\ d\theta\\
&-&   \int_0^{2\pi} t_m(\theta)^2(\frac{\partial \hat{\boldsymbol{m}}_t(\theta)}{\partial \theta} \cdot \hat{\boldsymbol{m}}_t(\theta)) d\theta. \nonumber \\
&+& \int_0^{2\pi} t_m(\theta)(\frac{\partial \hat{\boldsymbol{m}}_t(\theta)}{\partial \theta} \cdot \hat{\boldsymbol{T}}(\theta)) d\theta.
\end{eqnarray}
All terms above depends explicitly on $t_m(\theta)$, and thus if $\hat{\vec{\boldsymbol{M}}}_t(\theta)\cdot \hat{\boldsymbol{T}}(\theta)=0$ everywhere (e.g. the magnetization is out of the ring's plane only) there can be no resistance change.

We now observe that a term proportional to $\xi$, the degree of non-adiabaticity, the memristance change depends on $\frac{\partial \hat{\boldsymbol{m}}_t(\theta)}{\partial \theta}\cdot \Big(\hat{\boldsymbol{T}}(\theta)\times \hat{\boldsymbol{m}}_t(\theta) \Big)$, while the second term which is independent from $\xi$, on $(\frac{\partial \hat{\boldsymbol{m}}_t(\theta)}{\partial \theta} \cdot \hat{\boldsymbol{m}}_t(\theta))$ and $\frac{\partial \hat{\boldsymbol{m}}_t(\theta)}{\partial \theta} \cdot \hat{\boldsymbol{T}}(\theta)$. For systems which are fully magnetized, however, the second term is always zero. This can be seen by noticing that
\begin{eqnarray}
0=\partial_{\theta} 1=\partial_\theta \sum_i \hat m_i(\theta)^2=2\big( \partial_\theta \hat m_i(\theta)\big)\hat m_i(\theta)=2 \frac{\partial \hat{\boldsymbol{m}}_t(\theta)}{\partial \theta} \cdot \hat{\boldsymbol{m}}_t(\theta).
\end{eqnarray}

We now use another identity. We have

\begin{eqnarray}
t_m(\theta)(\frac{\partial \hat{\boldsymbol{m}}_t(\theta)}{\partial \theta} \cdot \hat{\boldsymbol{T}}(\theta))=\frac{1}{2}\partial_\theta t_m(\theta)^2-t_m(\theta)\big(\frac{\partial \hat{\boldsymbol{T}}(\theta)}{\partial \theta} \cdot \hat{\boldsymbol{m}}_t(\theta)\big)
\end{eqnarray}
Clearly, since $\frac{1}{2} \int_0^{2\pi} \partial_\theta t_m^2(\theta) d\theta=\frac{t_m(2\pi)^2-t_m(0)^2}{2}=0$ and, moreover, we have 
\begin{eqnarray}
\partial_\theta \hat{\boldsymbol{T}}(\theta)=-\hat{ \boldsymbol{R}}(\theta)
\end{eqnarray}
where $\hat{ \boldsymbol{R}}(\theta)=\big(\cos(\theta),\sin(\theta),0)$ is the vector oriented from the center of the  {ring}  towards the point at angle $\theta$, we have
\begin{eqnarray}
\int_0^{2\pi} t_m(\theta)(\frac{\partial \hat{\boldsymbol{m}}_t(\theta)}{\partial \theta} \cdot \hat{\boldsymbol{T}}(\theta)) d\theta&=&-\int_0^{2\pi} t_m(\theta)\big(\frac{\partial \hat{\boldsymbol{T}}(\theta)}{\partial \theta} \cdot \hat{\boldsymbol{m}}_t(\theta)\big) d\theta\nonumber \\
&=&\int_0^{2\pi} t_m(\theta)\big(\hat{\boldsymbol{R}}(\theta) \cdot  \hat{\boldsymbol{m}}_t(\theta)\big) d\theta
\end{eqnarray}
Let us call now $r_m(\theta)=\hat{\boldsymbol{R}}(\theta) \cdot \hat{\boldsymbol{m}}_t(\theta)$.

It follows that if the magnetization density is constant across the material, and changes in direction, then the second term can be neglected. It follows that in the approximation in which $\|\hat{\boldsymbol{m}}_t(\theta)\|^2=1$, then
\begin{eqnarray}
\frac{\beta(\hat{\boldsymbol{m}}_t)}{ \nu \Delta \rho }&=&- \xi \  \int_0^{2\pi} t_m(\theta)\Big( \frac{\partial \hat{\boldsymbol{m}}_t(\theta)}{\partial \theta}\cdot \Big(\hat{\boldsymbol{T}}(\theta)\times \hat{\boldsymbol{m}}_t(\theta) \Big)\Big)\ d\theta\\
&+& \int_0^{2\pi} t_m(\theta) r_m(\theta) d\theta\equiv \frac{\beta_{a}(\hat{\boldsymbol{m}}_t)}{ \nu \Delta \rho }+\frac{\beta_{na}(\hat{\boldsymbol{m}}_t)}{ \nu \Delta \rho }
\end{eqnarray}
where we implicitly defined the adiabatic and non-adiabatic quantities $\beta_a$ and $\beta_{na}$.

\subsection*{Supplementary Note 3}
The effective field with (micromagnetic) exchange interaction can be written as $\vec{\boldsymbol{H}}_\text{eff}=\vec{\boldsymbol{H}}_0+\lambda\partial^2_{\theta\theta}{\vec{\boldsymbol{M}}}_t(\theta)-H_{sh}(\hat{\boldsymbol{m}}_t(\theta)\cdot\hat{\boldsymbol{T}}(\theta))\hat{\boldsymbol{T}}(\theta)$, where $\lambda$ and $H_{sh}$ are effective parameters \cite{Iacocca2017}. Let us now replace $\vec{\boldsymbol{H}}_{eff}$ in $\eta(\hat{\boldsymbol{m}}_t,\vec{\boldsymbol{H}}_{eff})$. 

We have, since $a\cdot (a\times b)=0$, that
\begin{eqnarray}
\eta(\hat{\boldsymbol{m}}_{t})=r \Delta \rho \int_0^{2\pi} ( \hat{\boldsymbol{T}}(\theta)\cdot  \hat{\boldsymbol{m}}_t(\theta))((\gamma (\vec{\boldsymbol{H}}_0+\lambda M_S\partial^2_{\theta\theta}\hat{\boldsymbol{m}}_t(\theta))-\alpha \frac{\partial\hat{\boldsymbol{m}}_t(\theta)}{\partial t})\cdot \Big(\hat{\boldsymbol{T}}(\theta)\times \hat{\boldsymbol{m}}_t(\theta) \Big)\Big)\ d\theta
\end{eqnarray}

Let us now assume that $M_t(\theta)$ and $H_0$ are in plane. Then, $\vec{\boldsymbol{H}}_0\cdot (\hat{\boldsymbol{T}}(\theta)\times \hat{\boldsymbol{m}}_t)=0$, since $\hat{\boldsymbol{T}}\times \hat{\boldsymbol{m}}_t$ is out of plane. As a result, for highly biasing external fields, the coefficient $\eta$ is independent from the strength of $\vec{\boldsymbol{H}}_0$. We thus have

\begin{eqnarray}
\eta(\hat{\boldsymbol{m}}_t)=r \Delta \rho \int_0^{2\pi} ( \hat{\boldsymbol{T}}(\theta)\cdot  \hat{\boldsymbol{m}}_t(\theta))((\gamma (\vec{\boldsymbol{H}}_0+\lambda M_s\partial^2_{\theta\theta}\hat{\boldsymbol{m}}_t(\theta))-\alpha \frac{\partial\hat{\boldsymbol{m}}_t(\theta)}{\partial t})\cdot \Big(\hat{\boldsymbol{T}}(\theta)\times \hat{\boldsymbol{m}}_t(\theta) \Big)\Big)\ d\theta
\end{eqnarray}
If we instead assume that, up to a normalizing constant, $\hat{\boldsymbol{m}}_t\approx  \boldsymbol{P}_0+\epsilon \vec{\boldsymbol{Q}}_t(\theta)$, where we assume that $\boldsymbol{P}$ is in plane and $\vec{\boldsymbol{Q}}_t$ is a generic perturbation. We have

\begin{eqnarray}
\frac{\partial \tilde R}{\partial t}
&=&-\xi\nu \Delta \rho j(t)   \int_0^{2\pi} ( \hat{\boldsymbol{T}}(\theta)\cdot  (\boldsymbol{P}_0+\epsilon \vec{\boldsymbol{Q}}_t(\theta)))(\frac{\partial }{\partial \theta}(\vec{\boldsymbol{H}}_0+\epsilon \vec{\boldsymbol{Q}}_t(\theta))\cdot \Big(\hat{\boldsymbol{T}}(\theta)\times (\boldsymbol{P}_0+\epsilon \vec{\boldsymbol{Q}}_t(\theta)) \Big)\Big)\ d\theta \nonumber \\
&+& \nu  \Delta \rho\int_0^{2\pi} \big(\hat{\boldsymbol{T}}(\theta)\cdot (\boldsymbol{P}_0+\epsilon \vec{\boldsymbol{Q}}_t(\theta)) \big)\big(\hat{\boldsymbol{R}}(\theta)\cdot (\boldsymbol{P}_0+\epsilon \vec{\boldsymbol{Q}}_t(\theta)) \big)\nonumber \\
&-& r \Delta \rho \int_0^{2\pi} ( \hat{\boldsymbol{T}}(\theta)\cdot  \hat{\boldsymbol{m}}_t(\theta))((\gamma \vec{\boldsymbol{H}}_\text{eff}-\alpha \frac{\partial}{\partial t}(\boldsymbol{P}_0+\epsilon \vec{\boldsymbol{Q}}_t(\theta)))\cdot \Big(\hat{\boldsymbol{T}}(\theta)\times (\boldsymbol{P}_0+\epsilon \vec{\boldsymbol{Q}}_t(\theta)) \Big)\Big)\ d\theta \nonumber \\
&=&-\xi\nu \Delta \rho j(t)   \int_0^{2\pi} ( \hat{\boldsymbol{T}}(\theta)\cdot  \big(\boldsymbol{P}_0+\epsilon \vec{\boldsymbol{Q}}_t(\theta)\big))(\frac{\partial }{\partial \theta}(\epsilon \vec{\boldsymbol{Q}}_t(\theta))\cdot \Big(\hat{\boldsymbol{T}}(\theta)\times (\boldsymbol{P}_0+\epsilon \vec{\boldsymbol{Q}}_t(\theta)) \Big)\Big)\ d\theta \nonumber \\
&+& \nu  \Delta \rho\int_0^{2\pi} \big(\hat{\boldsymbol{T}}(\theta)\cdot (\boldsymbol{P}_0+\epsilon \vec{\boldsymbol{Q}}^{\|}_t(\theta)) \big)\big(\hat{\boldsymbol{R}}(\theta)\cdot (\boldsymbol{P}_0+\epsilon \vec{\boldsymbol{Q}}^{\|}_t(\theta)) \big)\nonumber \\
&-& r \Delta \rho \int_0^{2\pi} ( \hat{\boldsymbol{T}}(\theta)\cdot  (\vec{\boldsymbol{H}}_0+\epsilon \vec{\boldsymbol{Q}}_t(\theta)))((\gamma \vec{\boldsymbol{H}}_\text{eff}-\alpha \frac{\partial}{\partial t}(\epsilon \vec{\boldsymbol{Q}}_t(\theta)))\cdot \Big(\hat{\boldsymbol{T}}(\theta)\times (\boldsymbol{P}_0+\epsilon \vec{\boldsymbol{Q}}_t(\theta)) \Big)\Big)\ d\theta \nonumber \\
\end{eqnarray}
where we used the fact that $\vec{\boldsymbol{Q}}_t(\theta)\cdot \hat{\boldsymbol{R}}(\theta)=\vec{\boldsymbol{Q}}_t(\theta)\cdot \hat{\boldsymbol{T}}(\theta)=0$ if the perturbation is out-of-plane.
We now keep terms that are only up to the order $O(\epsilon^2)$.
\begin{eqnarray}
\frac{\partial \tilde R}{\partial t}
&=&- \epsilon\xi \nu j(t) \Delta \rho \int_0^{2\pi} ( \hat{\boldsymbol{T}}(\theta)\cdot  \boldsymbol{P}_0)(\frac{\partial }{\partial \theta}( \vec{\boldsymbol{Q}}_t(\theta))\cdot \Big(\hat{\boldsymbol{T}}(\theta)\times \boldsymbol{P}_0 \Big)\Big)\ d\theta \nonumber \\
&+& \nu  \Delta \rho\int_0^{2\pi} \big(\hat{\boldsymbol{T}}(\theta)\cdot (\boldsymbol{P}_0+\epsilon \vec{\boldsymbol{Q}}^{\|}_t(\theta)) \big)\big(\hat{\boldsymbol{R}}(\theta)\cdot (\boldsymbol{P}_0+\epsilon \vec{\boldsymbol{Q}}^{\|}_t(\theta)) \big)\nonumber \\
&-& r \Delta \rho \int_0^{2\pi} ( \hat{\boldsymbol{T}}(\theta)\cdot  (\boldsymbol{P}_0+\epsilon \vec{\boldsymbol{Q}}_t(\theta)))((\gamma \vec{\boldsymbol{H}}_\text{eff}-\alpha \frac{\partial}{\partial t}(\epsilon \vec{\boldsymbol{Q}}_t(\theta)))\cdot \Big(\hat{\boldsymbol{T}}(\theta)\times (\boldsymbol{P}_0+\epsilon \vec{\boldsymbol{Q}}_t(\theta)) \Big)\Big)\ d\theta \nonumber \\
\end{eqnarray}
We now replace $\vec{\boldsymbol{H}}_\text{eff}=\vec{\boldsymbol{H}}_0+\lambda  \partial^2_{\theta\theta}\vec{\boldsymbol{M}}_t=\vec{\boldsymbol{H}}_0+\lambda M_S \partial^2_{\theta\theta}\hat{\boldsymbol{m}}_t=\vec{\boldsymbol{H}}_0+\epsilon \lambda M_s\partial^2_{\theta\theta}\boldsymbol{Q_t}(\theta)$, where we assume that $\boldsymbol{H_0}$ is in plane. Then, if we define
 $\boldsymbol{G}=\gamma  \lambda M_s\partial^2_{\theta\theta}\boldsymbol{Q_t}(\theta)-\alpha \frac{\partial}{\partial t} \vec{\boldsymbol{Q}}_t(\theta)$, we have

\begin{eqnarray}
\frac{\partial \tilde R}{\partial t}
&=&-\xi \epsilon\nu j(t) \Delta \rho  \int_0^{2\pi} ( \hat{\boldsymbol{T}}(\theta)\cdot  \boldsymbol{P}_0)(\frac{\partial }{\partial \theta}( \vec{\boldsymbol{Q}}_t(\theta))\cdot \Big(\hat{\boldsymbol{T}}(\theta)\times \boldsymbol{P}_0 \Big)\Big)\ d\theta \nonumber \\
&+& \nu  \Delta \rho\int_0^{2\pi} \big(\hat{\boldsymbol{T}}(\theta)\cdot (\boldsymbol{P}_0+\epsilon \vec{\boldsymbol{Q}}^{\|}_t(\theta)) \big)\big(\hat{\boldsymbol{R}}(\theta)\cdot (\boldsymbol{P}_0+\epsilon \vec{\boldsymbol{Q}}^{\|}_t(\theta)) \big)\nonumber \\
&-& r \Delta \rho \int_0^{2\pi} \Big(\big( \hat{\boldsymbol{T}}(\theta)\cdot  (\vec{\boldsymbol{H}}_0+\epsilon \vec{\boldsymbol{Q}}_t(\theta))\big)(\gamma \boldsymbol{H_0}+\epsilon \boldsymbol{G} )\cdot \Big(\hat{\boldsymbol{T}}(\theta)\times (\boldsymbol{P}_0+\epsilon \vec{\boldsymbol{Q}}_t(\theta)) \Big)\Big)\ d\theta \nonumber \\
\end{eqnarray}
Let us focus on 
\begin{equation}
    (\gamma \vec{\boldsymbol{H}}_0+\gamma \epsilon \lambda M_s\partial^2_{\theta\theta}\boldsymbol{Q_t}(\theta)-\epsilon\alpha \frac{\partial}{\partial t} \vec{\boldsymbol{Q}}_t(\theta))\cdot \Big(\hat{\boldsymbol{T}}(\theta)\times (\boldsymbol{P}_0+\epsilon \vec{\boldsymbol{Q}}_t(\theta)) \Big)
\end{equation}
We can expand this term as,
\begin{eqnarray}
(\gamma \vec{\boldsymbol{H}}_0+\epsilon \boldsymbol{G})\cdot \Big(\hat{\boldsymbol{T}}(\theta)\times (\boldsymbol{P}_0+\epsilon \vec{\boldsymbol{Q}}_t(\theta)) \Big)
    &= &\cancel{\gamma \vec{\boldsymbol{H}}_0\cdot \Big(\hat{\boldsymbol{T}}(\theta)\times \boldsymbol{P}_0 \Big)}\nonumber \\
    &+ &\epsilon \gamma \vec{\boldsymbol{H}}_0\cdot \Big(\hat{\boldsymbol{T}}(\theta)\times  \vec{\boldsymbol{Q}}_t(\theta) \Big)\nonumber \\
    &+ &\epsilon \boldsymbol{G}\cdot \Big(\hat{\boldsymbol{T}}(\theta)\times \boldsymbol{P}_0 \Big)\nonumber \\
    &+ &\epsilon^2 \cancel{\boldsymbol{G}\cdot \Big(\hat{\boldsymbol{T}}(\theta)\times  \vec{\boldsymbol{Q}}_t(\theta) \Big)\nonumber} 
\end{eqnarray}
where the first term cancels because it is identically zero since both $\vec{\boldsymbol{H}}_0$ and $\boldsymbol{P}_0$ are assumed to be in plane, while the last  term is of order $\epsilon^2$. Since we assume that $\vec{\boldsymbol{H}}_0$ is on plane, it follows that these terms are non-zero only if $\vec{\boldsymbol{Q}}_t(\theta)$ has a component out-of-plane.
We then have
\begin{eqnarray}
\frac{\partial \tilde R}{\partial t}
&=&-\epsilon\nu \xi j(t) \Delta \rho  \int_0^{2\pi} ( \hat{\boldsymbol{T}}(\theta)\cdot  \boldsymbol{P}_0)(\frac{\partial }{\partial \theta}( \vec{\boldsymbol{Q}}^{\perp}_t(\theta))\cdot \Big(\hat{\boldsymbol{T}}(\theta)\times \boldsymbol{P}_0 \Big)\Big)\ d\theta \nonumber \\
&+& \nu  \Delta \rho\int_0^{2\pi} \big(\hat{\boldsymbol{T}}(\theta)\cdot (\boldsymbol{P}_0+\epsilon \vec{\boldsymbol{Q}}^{\|}_t(\theta)) \big)\big(\hat{\boldsymbol{R}}(\theta)\cdot (\boldsymbol{P}_0+\epsilon \vec{\boldsymbol{Q}}^{\|}_t(\theta)) \big)\nonumber \\
&-& \epsilon r \Delta \rho \int_0^{2\pi} ( \hat{\boldsymbol{T}}(\theta)\cdot  \boldsymbol{P}_0) \Big(\boldsymbol{G}\cdot \big(\hat{\boldsymbol{T}}(\theta)\times \boldsymbol{P}_0\big)+\gamma \vec{\boldsymbol{H}}_0\cdot \big(\hat{\boldsymbol{T}}(\theta)\times  \vec{\boldsymbol{Q}}^{\perp}_t(\theta) \big)\Big)\ d\theta \nonumber \\
\end{eqnarray}
where $\boldsymbol{G}=\gamma  \lambda M_s\partial^2_{\theta\theta}\boldsymbol{Q_t}(\theta)-\alpha \frac{\partial}{\partial t} \vec{\boldsymbol{Q}}_t(\theta)$.
We thus expect that, if $\boldsymbol{H_0}$ is strong enough to bias the magnetization $\hat{\boldsymbol{m}}_t$ along a certain direction in the same plane of $\boldsymbol{P}_{0}$, the change in the behavior of the resistance should not be consistent. For the normalization to be consistent, we need to choose $\epsilon\rightarrow \frac{\epsilon}{||\boldsymbol{P}_0||}$. 

As a result, we find that for $\beta_{a}$ and $\eta$, if both $\vec{\boldsymbol{H}}_0$ and $\boldsymbol{P}_0$ are in the same plane, the effect of $\vec{\boldsymbol{H}}_0$ is a smaller correction, and not of order $1$. For $\beta_{na}$ instead, only perturbations in plane are relevant. We thus predict that magnetomemristors are going to be robust to in plane perturbations.

\subsection*{Supplementary Note 4}
Let us note that the extra resistance is zero only if eqn. (\ref{eq:extra}) is equal to zero, e.g. if
\begin{eqnarray}
\frac{M_y(t,\theta)}{M_x(t,\theta)}=\tan(\theta).
\end{eqnarray}
Assume then that $M_x(\theta)=M_r(t,\theta) \cos(\theta)$ and $M_y(\theta)=M_r(t,\theta) \sin(\theta)$, and assume that $M^2=M_x^2+M_y^2+M_z^2$ is conserved. A first-order resistanceless state is thus given by
\begin{eqnarray}
M^2=M_{r}^2(t,\theta)+M_z^2(t,\theta),
\end{eqnarray}
from which it follows that $M_z=M\cos(f_t(\theta))$ and $M_r(\theta)=M\sin(f_t(\theta))$, 
for an arbitrary function of $f_t(\theta)$. This implies that there is an infinite family of magnetic configurations which has effectively zero magnetically induced resistance.
As an example, consider $M_z=0$. In this case we must have $M_x=M \sin(\theta)$ and $M_y=M \cos(\theta)$. These are states pointing towards or away from the center of the  {annulus}  at all $\theta$, e.g. radial magnetic states along the {annulus}.

\textit{Constant magnetic states} If $\hat{\boldsymbol{m}}_t(\theta)$ is constant instead, we have
\begin{eqnarray}
\hat{\boldsymbol{T}}(\theta)\cdot \hat{\boldsymbol{m}}_t(\theta)=M_y \cos\theta-M_x \sin(\theta).
\label{eq:extra2}
\end{eqnarray}
from which we obtain 
\begin{eqnarray}
\int_0^{2\pi}(\hat{\boldsymbol{T}}(\theta)\cdot \hat{\boldsymbol{m}}_t(\theta))^2\ d\theta=\pi  \left(M_x^2+M_y^2\right)=\pi (1-M_z^2)
\end{eqnarray}
assuming that the norm is conserved. We thus see here that if the magnetic field is perpendicular to the  {annulus}  plane, then the resistance does not have any AMR correction.
\subsection*{Supplementary Note 5}
We now study some magnetic states within the context of the  {annulus}  toy model, in which perfect magnetic states do not lead to memristance, in order to understand topological protection.
\subsubsection*{Non-adiabatic correction}
Let us first analyze the non-adiabatic correction. This is given by
\begin{eqnarray}
\beta_{na}(\hat{\boldsymbol{m}}_t)=\int_0^{2\pi} t_m(\theta)r_m(\theta)
\end{eqnarray}
It is easy to analyze the states for which the non-adiabatic term vanishes. These are those for which either $t_m(\theta)=0$ or for which $r_m(\theta)=0$.

These are given by magnetization states for which $\hat{\boldsymbol{m}}_t\perp \hat{\boldsymbol{T}}(\theta)$ or $\hat{\boldsymbol{m}}_t\perp \hat{\boldsymbol{R}}(\theta)$, which are respectively given by, setting $\hat {\boldsymbol{z}}=(0,0,1)$ and introducing the constants $a$, $b$ such that $a^2+b^2=1$,
\begin{eqnarray}
\hat{\boldsymbol{m}}_t^{\perp\hat{\boldsymbol{T}} }\propto a\hat{\boldsymbol{R}}(\theta)+b \hat{\boldsymbol{z}}, and\\
\hat{\boldsymbol{m}}_t^{\perp\hat{\boldsymbol{R}} }\propto a\hat{\boldsymbol{T}}(\theta)+b  \hat{\boldsymbol{z}}.
\end{eqnarray}
Since $t_m(\theta)$ is also contained in $\eta$ and in the adiabatic correction, it follows that states of the form $\hat{\boldsymbol{m}}_t^{\perp\hat{\boldsymbol{T}} }$ are stable memristanceless states. We stress that this is true only for the case of a  {annulus}  of zero thickness, and in the absence of dipole interactions. Indeed, as we show in the main text, the interplay between magnetic kinks and dipoles are the origin of the memristance in the realistic finite size simulations.

{
\textit{Non-adiabatic maximal and minimal states.} 
There are however some states which present a maximum value of $\beta$. In particular,  $\beta_{na}$  would be dominating in several experiments, given that typically $\xi\ll 1$.  For this purpose, we need to maximize $\hat{\boldsymbol{m}}^*=\text{max}_{\hat{\boldsymbol{m}}}\beta_{na}(\hat{\boldsymbol{m}})$.
It is not hard to see that the maximum value that both $t_m(\theta), r_m(\theta)$ can have is $1/2$ if $\hat{\boldsymbol{m}}_t(\theta)=\pm\frac{1}{\sqrt{2}} [\hat{\boldsymbol{R}}(\theta)+\hat{\boldsymbol{T}}(\theta)]$ (non-adiabatic maximum state), while the minimum value is $-\frac{1}{2}$ if $\hat{\boldsymbol{m}}_t(\theta)=\pm \frac{1}{\sqrt{2}} [\hat{\boldsymbol{R}}(\theta)-\hat{\boldsymbol{T}}(\theta)]$ (non-adiabatic minimum state). Since $\hat{\boldsymbol{R}}(\theta)\cdot\hat{\boldsymbol{T}}(\theta)=0$, we have in both cases $\|\hat{\boldsymbol{m}}_t\|^2=1$. These states, shown in Supplementary Figure \ref{fig:nonadmag}, correspond to a maximal rate of change of the resistivity due to the non-adiabatic interaction, for a non-zero current.
We thus obtain a reference value for $\xi$: if $\xi\gg \frac{1}{2}$, then the non-adiabatic term dominates, while if $\xi\ll \frac{1}{2}$ the adiabatic term dominates. Such static states would be difficult to stabilize in a trivial ferromagnet.
\begin{figure}
    \centering
    \includegraphics[scale=0.3]{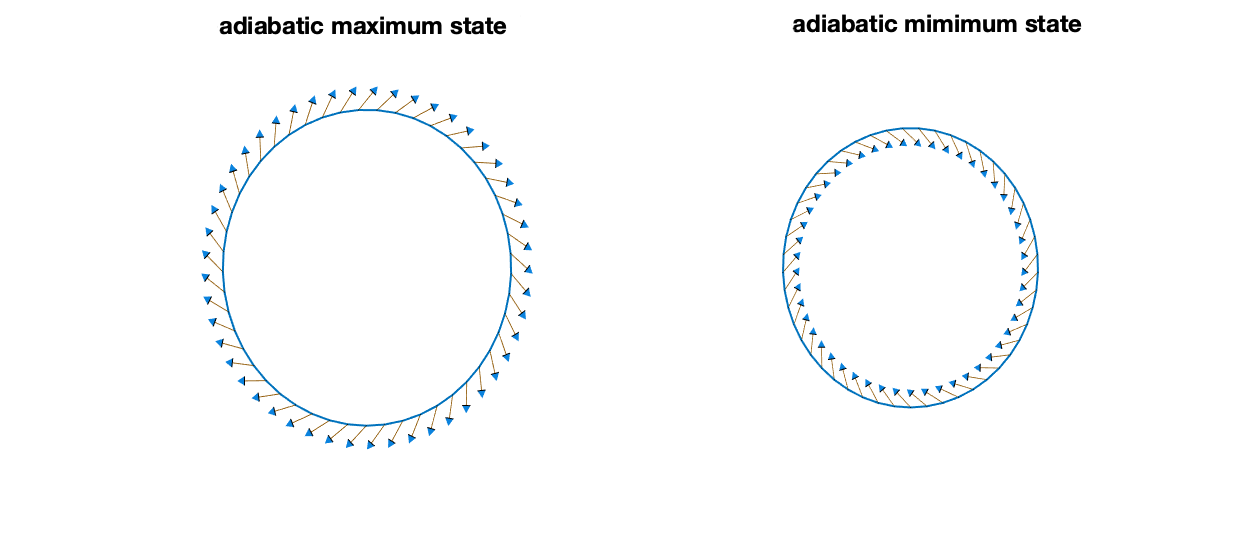}
    \caption{\textbf{Adiabatic minimum and maximum effects.} he two magnetic states from which a maximum absolute change in resistivity is attained in the non-adiabatic term of $\beta$.}
    \label{fig:nonadmag}
\end{figure}
}
\subsubsection*{Adiabatic correction}
Let us consider the constant $\beta$ for certain particular helical states of the magnetization along the ring.
\begin{eqnarray}
\beta_a(\hat{\boldsymbol{m}}_t)
&=&\nu \Delta \rho  \int_0^{2\pi} ( \hat{\boldsymbol{T}}(\theta)\cdot  \hat{\boldsymbol{m}}_t(\theta))(\frac{\partial \hat{\boldsymbol{m}}_t(\theta)}{\partial \theta}\cdot \Big(\hat{\boldsymbol{T}}(\theta)\times \hat{\boldsymbol{m}}_t(\theta) \Big)\Big)\ d\theta \nonumber \\
\end{eqnarray}
We recall that
\begin{eqnarray}
\bold{T}(\theta)=[-\sin \theta,\cos \theta,0]. 
\end{eqnarray}
We note that
\begin{eqnarray}
\hat{\boldsymbol{T}}(\theta)\times \hat{\boldsymbol{m}}_t(\theta)=\Big(T_y M_z-T_z M_y, T_z M_x-T_x M_z,T_x M_y-T_y M_x \Big).
\end{eqnarray}
\subsubsection*{x-y walls}
First, we consider a magnetic configuration which corresponds to $N$ planar Ne\'{e}l walls. The magnetic state is given by
\begin{eqnarray}
\hat{\boldsymbol{m}}_t(\theta)=\Big(\cos(\omega \theta),\sin(\omega \theta) ,0 \Big)
\end{eqnarray}
where we impose that $\hat{\boldsymbol{m}}_t(0)=\hat{\boldsymbol{m}}_t(2\pi)$. Thus, we can write $\omega = N$ up to translations in the $\theta$ direction. We have
\begin{eqnarray}
\hat{\boldsymbol{T}}(\theta)\cdot \hat{\boldsymbol{m}}_t(\theta)&=&-\sin(\theta)\cos(\omega \theta) +\sin(\omega\theta) \cos(\theta)\\
\partial_\theta \hat{\boldsymbol{m}}_t(\theta)&=&\omega \Big(-\sin(\omega \theta),\cos(\omega \theta) ,0 \Big)\\
\hat{\boldsymbol{T}}(\theta)\times \hat{\boldsymbol{m}}_t(\theta)&=&-\Big(0,0,\cos(\theta)\cos(\omega \theta)+\sin(\theta)\sin(\omega \theta)\Big)
\end{eqnarray}
It follows that $\partial_\theta \hat{\bold{m}}_t(\theta) \cdot (\hat{\boldsymbol{T}}(\theta)\times \hat{\boldsymbol{m}}_t(\theta))=0$, and thus $\beta_{a}(\hat{\bold{m}}_t)=0$ identically for any $\omega$.
\subsubsection*{y-z walls}

Let us consider now a Ne\'{e}l wall in the y-z direction. 

The magnetic state is given by
\begin{eqnarray}
\hat{\boldsymbol{m}}_t(\theta)=\Big(0,\cos(\omega \theta),\sin(\omega \theta) \Big)
\end{eqnarray}
where we impose again that $\hat{\boldsymbol{m}}_t(0)=\hat{\boldsymbol{m}}_t(2\pi)$. We have
\begin{eqnarray}
\hat{\boldsymbol{T}}(\theta)\cdot \hat{\boldsymbol{m}}_t(\theta)&=&\cos(\omega \theta) \cos(\theta)\\
\partial_\theta \hat{\boldsymbol{m}}_t(\theta)&=&\omega \Big(0,-\sin(\omega \theta),\cos(\omega \theta)  \Big)\\
\hat{\boldsymbol{T}}(\theta)\times \hat{\boldsymbol{m}}_t(\theta)&=&\Big(\cos(\omega \theta)\cos(\theta),0,-\sin(\theta)\cos(\omega \theta)\Big).
\end{eqnarray}
It follows that $\partial_\theta \hat{\bold{m}}_t(\theta) \cdot (\hat{\boldsymbol{T}}(\theta)\times \hat{\boldsymbol{m}}_t(\theta))=0$, and thus $\beta_{na}(\hat{\bold{m}}_t)=0$ identically for any $\omega$.
Thus,  
\begin{eqnarray}
\partial_\theta \hat{\boldsymbol{m}}_t(\theta)\cdot \Big(\hat{\boldsymbol{T}}(\theta)\times \hat{\boldsymbol{m}}_t(\theta)\Big)=-\sin(\theta)\cos^2(\omega \theta).
\end{eqnarray}
We have
\begin{eqnarray}
\beta_{a}(\hat{\boldsymbol{m}}_t)
&=&-\nu \Delta \rho \int_0^{2\pi} \cos^3(\omega \theta) \sin(\theta)\cos(\theta)d\theta =0,\ \forall \omega\nonumber \\
\end{eqnarray}

\subsubsection*{Helical states.} Next, we consider a helical magnetic state along the ring, given by
\begin{eqnarray}
\hat{\boldsymbol{m}}_t(\theta)=(M_x,M_y,M_z)&=& m_0\Big(\sin(\phi_1(\theta))\cos ( \phi_2(\theta)),\nonumber 
 \sin( \phi_1(\theta)) \sin(\phi_2(\theta)),
 \cos(\phi_1(\theta)) \Big)
\end{eqnarray}
where we require that the functions $\phi_1=\omega_1 \theta$ and $\phi_2=\omega_2\theta$ to be linear, with $\hat{\boldsymbol{m}}_t(0)=\hat{\boldsymbol{m}}_t(2\pi)$, up to some translation in the $\theta$ direction. It follows that $\omega_1 2\pi= n 2\pi$, and $\omega_2 2\pi= k 2\pi$ for integers $n$ and $k$. 
It is not hard to see that $\hat{\boldsymbol{m}}_t(\theta)\cdot \hat{\boldsymbol{T}}(\theta)=0$ everywhere. This implies that the resistivity reduces %ance the resistance reduced 
to $\rho_0$, the material natural resistivity. This is a special case of the memristanceless case. Since $\beta$'s and $\eta's$ also depend on $t_m(\theta)$, these states also lead to $\partial_t \tilde{\rho}=0$. Examples of these states are shown in Supplementary Figure \ref{fig:prot}.

\begin{figure}[ht!]
    \centering
    \includegraphics[width=\textwidth]{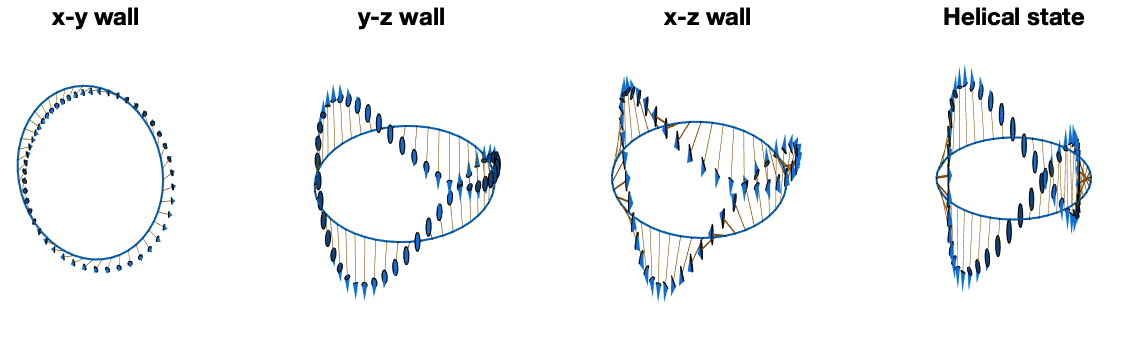}
    \caption{\textbf{Topologically nontrivial magnetic states.} Magnetic states in which the resistivity $\rho_0$ is protected from the magnetomemristive effect.}
    \label{fig:prot}
\end{figure}
\subsubsection*{Perfect spin superflows}
For spin superflows we have 
\begin{equation}
    \hat{{\boldsymbol{m}}}(\theta,\phi)= \sin(\phi) \cos(n\theta) \hat x+\sin(\phi) \sin(n\theta) \hat y+\cos(\phi) \hat z.
\end{equation}
We have
\begin{eqnarray}
t_m(\theta)&=&\hat{\boldsymbol{T}}(\theta)\cdot  \hat {\boldsymbol{m}}=-\sin(\phi)\cos(n\theta) \sin(\theta)+\sin(\phi)\cos(\theta) \sin(n\theta)\\
r_m(\theta)&=&\hat{\boldsymbol{R}}(\theta)\cdot  \hat {\boldsymbol{m}}=\cos(\theta) \cos(n \theta)\sin(\phi)+\sin(\theta)\sin(\phi)\sin(n\theta).
\end{eqnarray}
These states,  do contribute to the resistivity. We have in fact that 
\begin{eqnarray}
\rho=\rho_0+ \frac{\Delta \rho}{2\pi} \int_0^{2\pi} d\theta\  t_m^2(\theta)=\rho_0+\frac{\Delta \rho}{2\pi} \left(\pi -\frac{\sin (4 \pi  n)}{4 (n-1)}\right) \sin ^2(\phi )
\end{eqnarray}
Clearly, for $n=1$ the latter term in equation above is zero, as $t_m=0$ in that case. For any other integer $n$ however, we have
\begin{eqnarray}
\rho=\rho_0+ \frac{\Delta \rho}{2\pi} \int_0^{2\pi} d\theta\  t_m^2(\theta)=\rho_0+\frac{\Delta \rho}{2}  \sin^2(\phi)
\end{eqnarray}
which we see depends explicitly on the angle $\phi$.
We now test whether $\beta$'s are non zero, in which case we have a memristive effect.

From which we can perform the integral analytically for $\beta_{a}$ for the resistivity change, obtaining 

\begin{eqnarray}
\beta_{a}(\hat{\boldsymbol{m}}_t)
&=&-\nu \Delta \rho  \frac{1}{2\pi}\int_0^{2\pi} \cos^3(\omega \theta) \sin(\theta)\cos(\theta)d\theta =0,\ \forall \omega\nonumber \\
&=&\nu \Delta \rho \frac{\sin ^2(2 \pi  n) \sin ^2(\phi )}{4 \pi  (n-1)}
\end{eqnarray}
from which we see that if $n$ is integer, we have $\beta_{a}(\hat{\boldsymbol{m}}_t)=0$. This implies that while we have an AMR effect contributing to a static resistance, there is no memristive effect.

\subsubsection*{Localized kink}
We now discuss a localized magnetic kink, which is a model for the {2DW} state.

This state is sum of two magnetization vectors, 
\begin{eqnarray}
\hat{\boldsymbol{m}}(\theta)=a(\theta) \hat{\boldsymbol{T}}(\theta)+b(\theta) \hat{\boldsymbol{K}}(\theta)
\end{eqnarray}
where we assume $\boldsymbol{K}\cdot \hat{\boldsymbol{T}}=0$, and thus $|\hat{\boldsymbol{m}}|^2=a(\theta)^2+b(\theta)^2=1$.

We assume, in particular, that $a(\theta)$ and $b(\theta)$ are non-zero over almost non-overlapping regions, meaning that the magnetic state transitions over a localized region from the ground state (which is $\hat{\boldsymbol{T}}$) to $\boldsymbol{K}$, which is why we call this state a localized kink. A state formally of this type is obtained from the relaxation of domain walls in the extended system in the main text.

It is not hard to see that the resistive state is connected to $t_m(\theta)= a(\theta)$, and
\begin{eqnarray}
\tilde \rho=\rho_0+\frac{\Delta \rho}{2\pi } \int_0^{2\pi} d\theta\ a(\theta)^2.
\end{eqnarray}

Then, we assume that $r_m(\theta)=b(\theta)\hat{\boldsymbol{R}}(\theta)\cdot \hat{\boldsymbol{K}}(\theta)$ is non-zero. Then
\begin{eqnarray}
\beta_{a}=-\frac{\nu \Delta \rho}{2\pi} \int_0^{2\pi} a(\theta) b(\theta) \hat{\boldsymbol{R}}(\theta)\cdot \hat{\boldsymbol{K}}(\theta)\ d\theta.
\end{eqnarray}
From the equation above, we see that if the transition from $a(\theta)=1$ to $b(\theta)=1$ is smooth, as long as $\hat{\boldsymbol{R}}(\theta)\cdot \hat{\boldsymbol{K}}(\theta)\neq 0$, then there can be a memristive behavior. This is in essence why {the 2DW} state can have both AM resistance and memristance, unlike a perfect spin superfluid.

\section*{Supplementary Note 6}
Below we provide further details on numerical simulations.
%\subsection*{Supplementary Note 2.1}
It is understood that the high aspect ratio of the micromagnetic cells can significantly affect the results when the magnetic textures are potentially distorted. For low current densities, when the domain walls are approximately rigid, the qualitative behavior remains unchanged. For example, we reproduce the AMM in a  {annulus}  of thickness $5$~nm (as opposed to $10$~nm in the main text), a current density of $10^{11}$~A~m$^{-2}$, and a frequency $f=1$~GHz. The resulting Lissajous curves are shown in Supplementary Figure \ref{fig:thickness}, where we immediately observe the AMM effect. The quantitative difference in the curves may be associated in part with the cell's aspect ratio but we must stress that the reduced magnetic volume will also decrease the magnitude of the non-local dipole and, in turn, this will lead to a reduced out-of-plane magnetization components in the domain wall (helical distortion) %of the SS state 
that ultimately gives rise to memristance.
\begin{figure}
    \centering
    \includegraphics[scale=0.5]{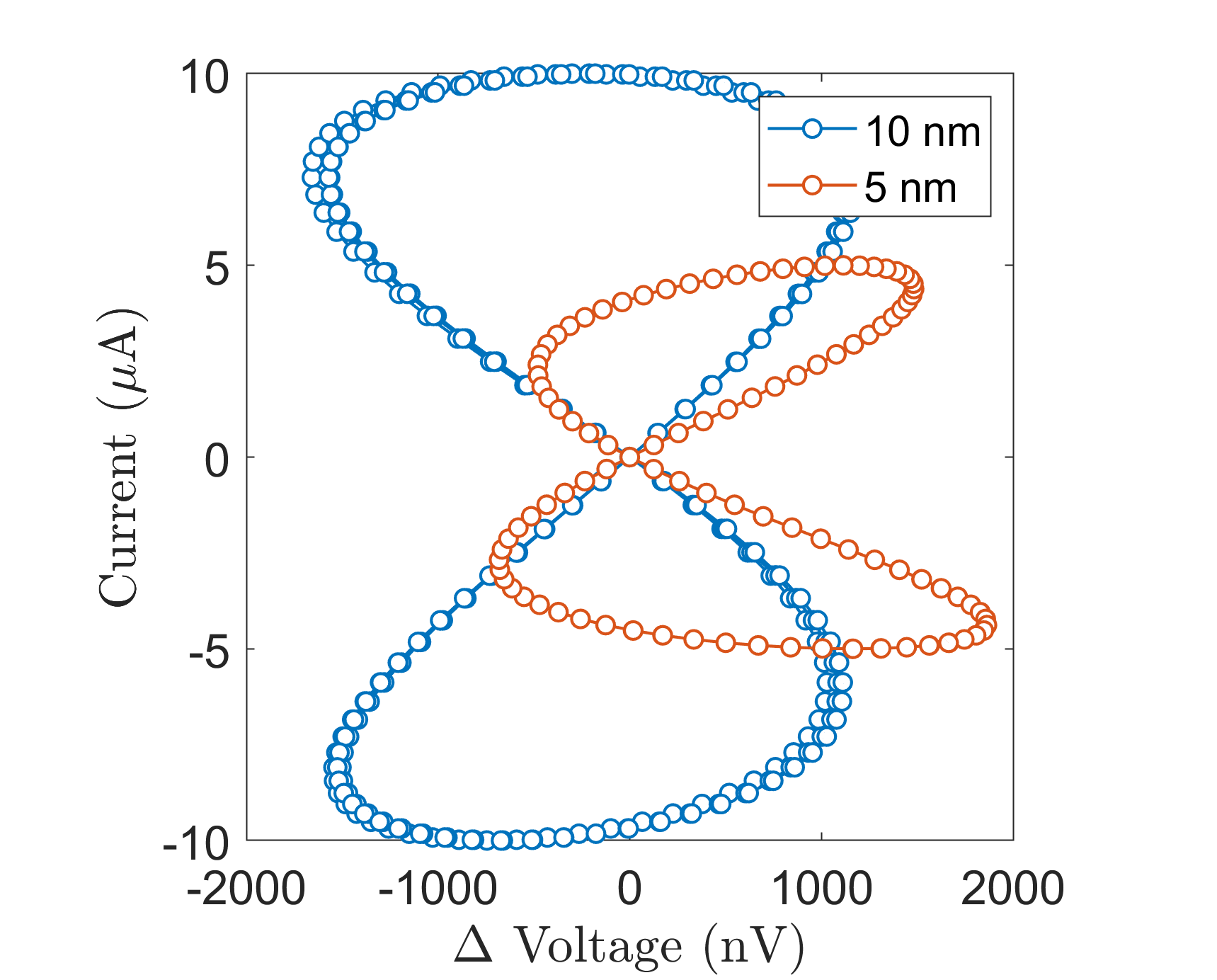}
    \caption{\textbf{Hysteresis and thickness.} Lissajous curves for rings with $10$~nm and $5$~nm thickness. The current in the $5$~nm is smaller because of the reduced cross-area by a factor $2$. Despite this difference, the maximum voltage has a similar magnitude.}
    \label{fig:thickness}
\end{figure}

\section*{Supplementary Note 7}

The  {annulus}  radius can play an important role in the memristance, as seen by the inverse proportionality of $\beta$ in Eqs.~\eqref{eq:memlike_full}. While exploring the effect of varying $r$ can be considered to be a project on its own, initial simulations suggest that indeed the inverse proportionality trend holds. We maintained the  {annulus}  width to $10$~nm throughout the simulations and modified the average  {annulus}  radius $r$. For $r=45$~nm the largest change in $d\tilde{\rho}/dt$ was $22.2$~k$\Omega$~s$^{-1}$ while for $r=300$~nm we obtained a change of $0.58$~k$\Omega$~s$^{-1}$. However, the Lissajous area in the I-V characteristics appeared to vary only slightly on the order of $1.6$~$\mu$W. This is encouraging for experiments where the effect could be measured for much larger rings that can be realistically patterned with traditional lithography techniques.

\end{document}